\documentclass[usenatbib]{mn2e}

\usepackage{graphicx}
\usepackage{lscape}
\usepackage{rotating}
\usepackage{scalefnt}

\newif\ifAMStwofonts

\bibpunct{(}{)}{;}{a}{}{,}

\def\eg{{\rm e.g. }}

\def\ie{{\rm i.e. }}

\newcommand{\kms}{km~s$^{-1}$\,}

\title[Analysis and calibration of CaII triplet spectroscopy] 
{Analysis and calibration of CaII triplet spectroscopy of Red Giant Branch stars from VLT/FLAMES observations}
\author[G. Battaglia et al.]  {G.Battaglia$^1$\thanks{Corresponding author.
E-mail:gbattagl@astro.rug.nl}, M.Irwin$^2$, E.Tolstoy$^1$, V.Hill$^3$, A.Helmi$^1$, 
B.Letarte$^{1,4}$, P.Jablonka$^5$\\ \\ $^1$Kapteyn Astronomical 
Institute, University of Groningen, P.O.Box 800, 9700 AV
Groningen, The Netherlands\\ $^2$ Institute of Astronomy, Madingley Road, Cambridge CB3 0HA  
\\ $^3$ GEPI, Observatoire de Paris, CNRS, Université Paris Diderot ; Place Jules Janssen 92190 Meudon, France\\ 
$^4$ Caltech Astronomy, 
MC105-24, Pasadena, CA 91125, USA\\ 
 $^5$ Observatoire de Gen\`eve, Laboratoire d'Astrophysique de
l'Ecole Polytechnique F\'ed\'erale de Lausanne (EPFL), \\ 
CH-1290 Sauverny,
Switzerland}

\pagerange{\pageref{firstpage}--\pageref{lastpage}}
\pubyear{}

\begin{document}

\maketitle

\label{firstpage}

\begin{abstract}
We demonstrate that low resolution Ca\,{\small II} triplet (CaT) spectroscopic
estimates of the overall metallicity ([Fe/H]) of individual Red Giant
Branch (RGB) stars in two nearby dwarf spheroidal galaxies (dSphs) agree to
$\pm$0.1-0.2 dex with detailed high resolution spectroscopic determinations
for the same stars over the range $-2.5 <$ [Fe/H] $< -0.5$. 
For this study we used a sample of 129 stars observed in 
low and high resolution mode with VLT/FLAMES in the Sculptor and 
Fornax dSphs. We also present the data reduction steps we used 
in our low resolution analysis and show that the typical 
accuracy of our velocity and CaT [Fe/H]  
measurement is $\sim$2 \kms and 0.1 dex respectively. 
We conclude that CaT-[Fe/H] relations calibrated on globular clusters can be 
applied with confidence to RGB stars in composite stellar 
populations over the range $-2.5 <$ [Fe/H] $< -0.5$. 
\end{abstract}


\section{Introduction}
An important aspect for a full understanding of galactic evolution is the
metallicity distribution function of the stellar population with time.  

Carrying out detailed abundance analyses with high resolution (HR)
spectroscopy to trace the patterns that allow one to distinguish
between the different galactic chemical enrichment processes is time
consuming for large samples of individual stars in a galaxy. This is
partly due to the observing time required, but also because of the
complex data reduction and analysis necessary.  Fortunately, there is
an empirically developed, simply calibrated method available which can
make an efficient estimate of metallicity ([Fe/H]) for individual Red
Giant Branch (RGB) stars using the strength of the Ca\,{\small II} triplet (CaT)
lines at 8498, 8542, 8662 \AA. This method was pioneered for use on
individual stars by \citet{arm1991}. It has the advantage that the
lines are broad enough that they can be accurately measured with
moderate spectral resolution \citep[e.g,][]{cole2004}.

The CaT method is routinely used to estimate [Fe/H] for nearby
resolved stellar systems and also provides an accurate radial velocity
estimate.  Both measurements are facilitated by the strength of the
CaT lines and by the generally red colours of the target stars.
However, the CaT-derived abundances are empirically defined, with a
poorly understood physical basis. Therefore it is important to check
the results against HR spectroscopic (\ie direct) measurements of
[Fe/H] and other elements.  The ``classical'' CaT calibration is based
on the use of globular cluster stars, all which are drawn from a
single age and metallicity stellar population.  The CaT equivalent
widths are directly compared to HR spectroscopic measurements of
[Fe/H] over a range of metallicity, and this comparison is used to
define the relation between CaT equivalent width (EW) and [Fe/H] for
all observations taken with the same set up.  This approach has been
extensively tested for a large sample of globular clusters
\citep[and references therein]{rutledge1997a, rutledge1997b}. 
However, globular clusters
typically exhibit a constant [Ca/Fe] for a large range of [Fe/H]. This
leads to uncertainty in the effect of varying [Ca/Fe] ratios such as
is seen in the more complex stellar populations found in galaxies.
Furthermore, stars in dwarf galaxies invariably cover a significant
range of ages as well as metallicities.  This mismatch in the
properties of calibrators and targets has led to suggestions that the
CaT method may not be a very accurate indicator of [Fe/H] for more
complex stellar populations, especially in those cases where [Ca/Fe]
varies significantly \citep[e.g,][]{pont2004}.

In this paper we investigate the validity of the CaT method for
complex stellar populations. We compare large samples of [Fe/H]
measurements coming from VLT/FLAMES made using both the CaT method and
direct HR
 spectroscopic measurements {\it for the same stars} in two
nearby dwarf spheroidal galaxies (dSphs), Sculptor and Fornax, over a
range of [Fe/H] and [Ca/H]. This is the first time such a detailed
comparison has been made for stars outside globular clusters.  We also
investigate the theoretically predicted behaviour of the CaT method
for a range of stellar atmospheric parameters using a grid of model
atmosphere spectra from \citet{munari2005}.

The paper is organised as follows.  In Section~\ref{sect:datareduction} we describe the data
reduction steps we use within the DART (Dwarf galaxy Abundances and
Radial velocities Team) 
collaboration to estimate EW
and velocities from observations in the CaT region, as the accuracy
with which this can be done clearly has important implications for the
reliability of our conclusions for these galaxies.  We also discuss
the verification of the overall calibration and accuracy of the
velocity and EW measurements by comparison of results from independent
CaT observations and by comparing with theoretical expectations based
on signal-to-noise, resolution and line profile properties. 
In Section~\ref{sec:calibration_gcs} we derive the standard CaT-[Fe/H] globular cluster
calibration for low resolution (LR) VLT/FLAMES data. In Section~\ref{sec:calibration_hr} we
compare the derived [Fe/H] from the CaT to the HR [Fe/H] for the Sculptor and Fornax dSphs.
 Finally, in Section~\ref{sec:uncert} we discuss the uncertainties that come
from using Ca\,{\small II} lines to derive an [Fe/H] abundance for stellar
populations where the $\alpha$-abundance varies and use a comparison
with stellar model atmospheres to further investigate age, metallicity
and $\alpha$-abundance effects.

\section{Low Resolution Data reduction \& Analysis} \label{sect:datareduction}

\begin{figure}
\begin{center}
\includegraphics[width=70mm]{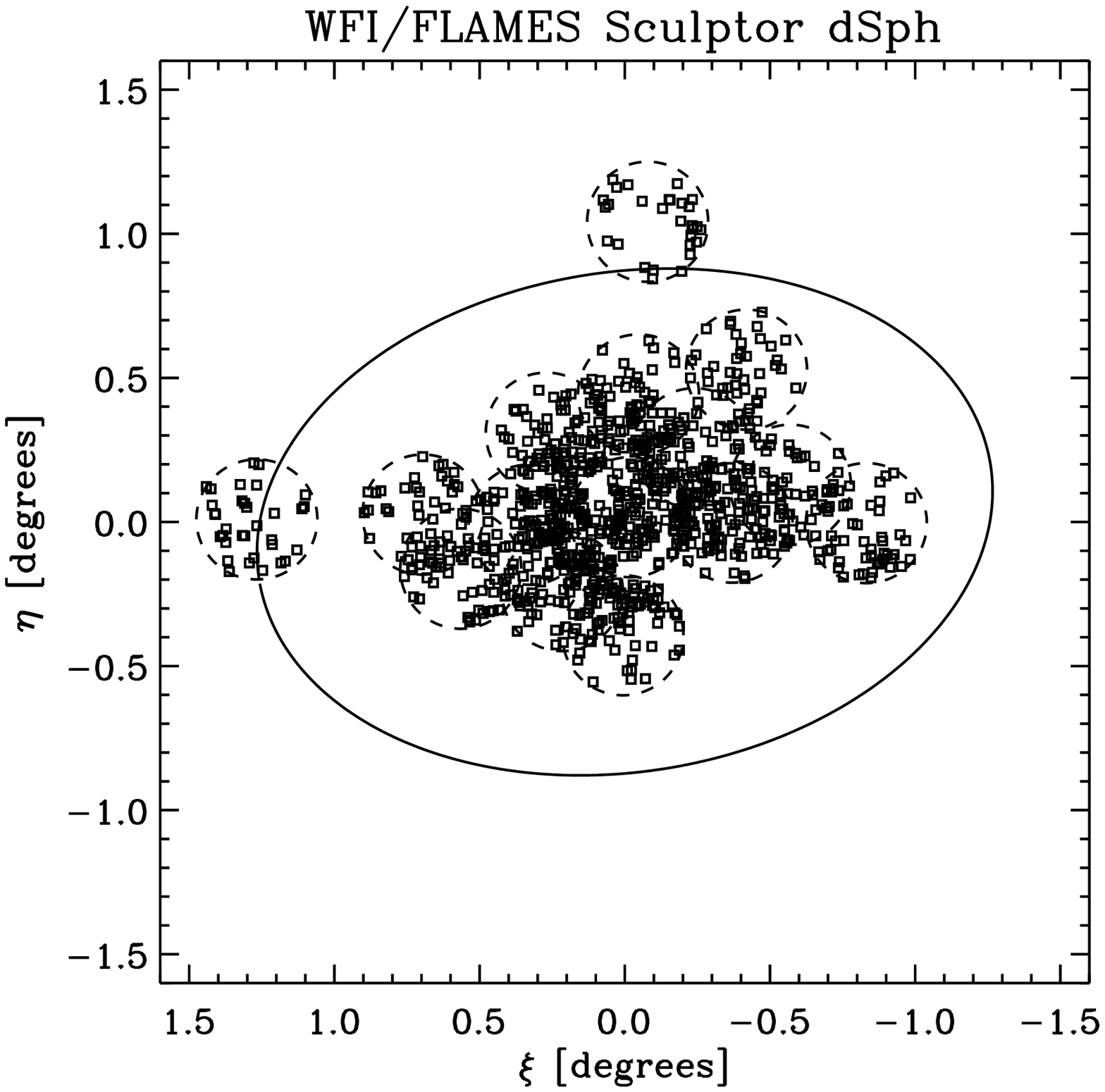}
\includegraphics[width=70mm]{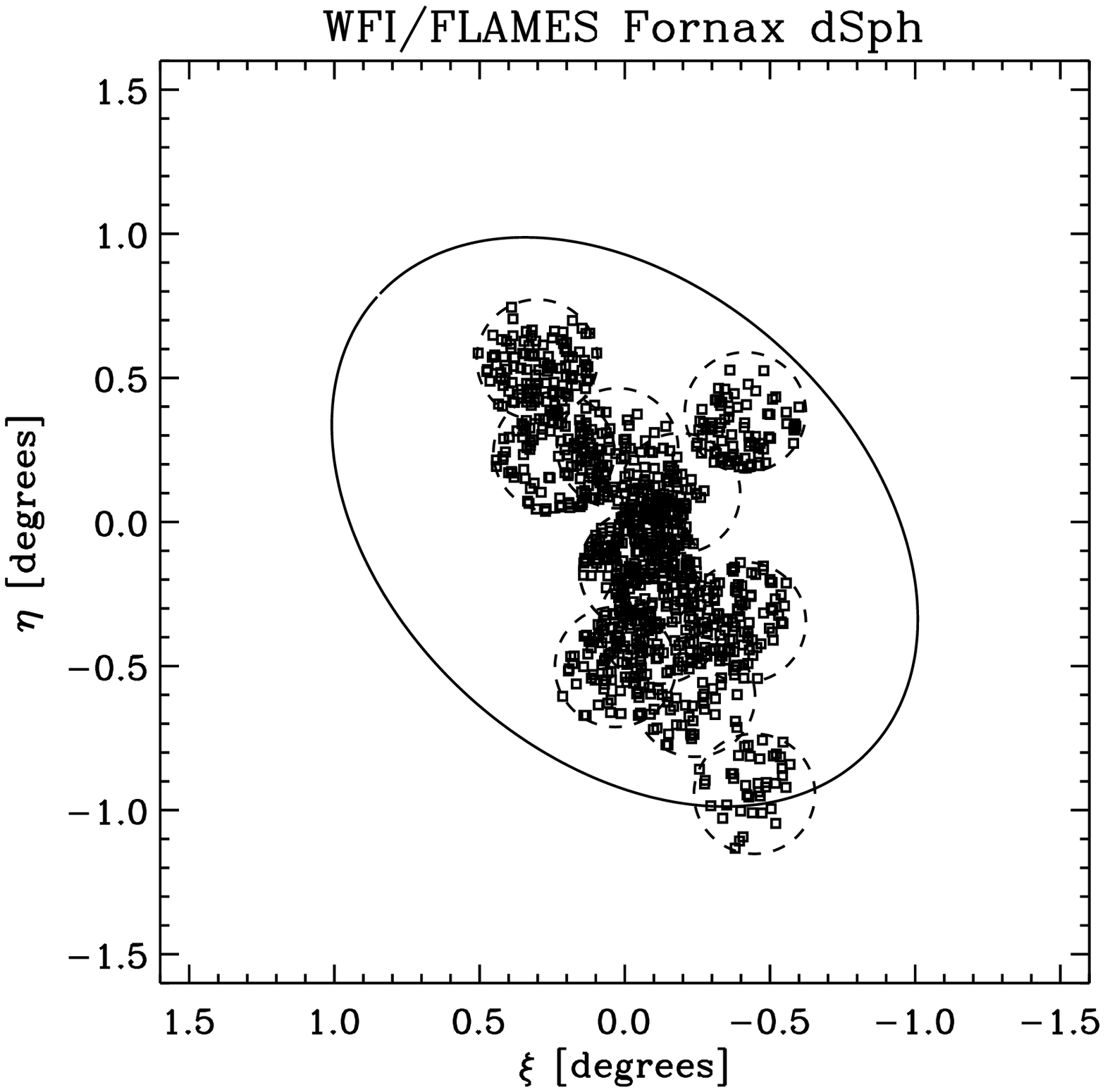}
\caption{Locations of the observed FLAMES targets at LR (squares) for the Sculptor dSph (top), 
centred at $01^h 00^m 09^s$ $-33^{\circ} 42' 30''$ \citep{Mateo1998} with position 
angle 99$^{\circ}$ \citep{IH1995}, and for the 
Fornax dSph (bottom), centred at $02^h 39^m 52^s$ $-34^{\circ} 30' 49''$ 
with position angle 46.8$^{\circ}$ \citep{battaglia2006}. 
The ellipses indicate the tidal 
radius (Sculptor: Irwin \& Hatzidimitriou \citeyear{IH1995}; Fornax: Battaglia et al. \citeyear{battaglia2006}). 
The dashed circles show the FLAMES pointings. North is up and East is to the left.}
\label{fig:fov}
\end{center}
\end{figure}

The datasets presented here were collected between August 2003 and
November 2005. They consist of 15 pointings in the Sculptor dSph and 11 in the Fornax
dSph spread over the galaxies (Fig.~\ref{fig:fov}).  Some fields were
observed with 1 hour exposure time, whilst other fields have repeated
exposures of shorter integration time and two different plate set-ups,
with the aim of testing the reliability of the derived velocities, EWs
and the stability of the instrument. 

All the LR CaT observations were made using VLT/FLAMES in Medusa
mode. This allows the simultaneous allocation of up to 132 fibres,
including dedicated sky fibres, over a 25' diameter field-of-view.  We
used the GIRAFFE low resolution grating (LR8), which covers the
wavelength range from 8206\AA\,
-- 9400\AA, and gives a resolution of R$\approx$6500. This allows the
measurement of EW from CaT lines and also enables the derivation of
velocities accurate to a $\approx$few km/s.  This set-up was used as
part of the DART programme to obtain spectra for several different
fields in each of the Sculptor, Fornax and Sextans dSphs, and also
for calibration purposes on a sample of 4 globular clusters: NGC104,
NGC5904, NGC3201, NGC4590\footnote{We acquired the data for 
NGC104 during an observing run in January 2005, whilst the data 
for the other globular clusters 
are from the ESO archive.}, which cover the range $-2.0 \la $ [Fe/H]
$\la -0.7$ on the CG97 scale (see Table~\ref{tab:gc}).

The central fields in Sculptor and Fornax were also observed with a
similar VLT/FLAMES set-up but at HR with R$\approx$20000, which
facilitates direct measurement of individual lines, and hence direct
abundance determination, of numerous elements (see Sect.~\ref{sec:calibration_hr}). 

Table~\ref{tab:journal} shows the journal of the VLT/FLAMES LR and HR observations we used for our 
analysis of the CaT-[Fe/H] calibration (Sects.~\ref{sec:calibration_gcs}, \ref{sec:calibration_hr}).

The data were all initially reduced using the GIRBLDRS\footnote{available
at SourceForge, http://girbldrs.sourceforg.net/} pipeline provided by the 
FLAMES consortium \citep[Geneva Observatory,][]{blecha2003}.  This package
provided flat-fielding (including fringing removal),
individual spectral extraction and accurate wavelength calibration,
based on daytime calibration exposures.  
At the time we started this project no sky subtraction was available within 
this pipeline which led us to develop several further reduction stages for
the LR analysis. We describe the LR analysis in detail here. In Tables~\ref{tab:results_gcs} and \ref{tab:results_dsphs}
we present the LR and HR results relevant for our analysis of the CaT-[Fe/H] 
calibration (Sects.~\ref{sec:calibration_gcs}, \ref{sec:calibration_hr}, \ref{sec:uncert}).

\begin{figure*}
\begin{center}
\includegraphics[width=120mm,angle=270]{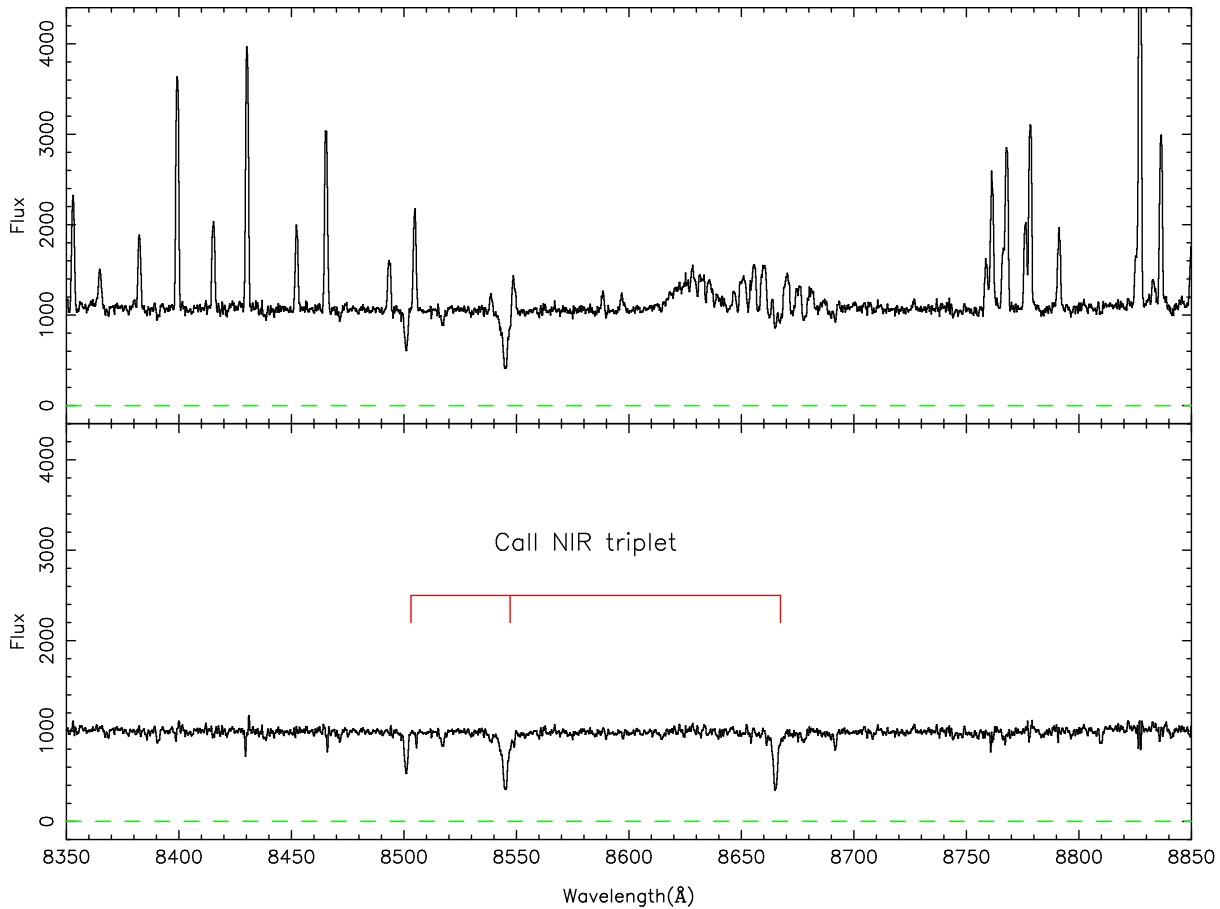}
\caption{An example of a LR spectrum from FLAMES in the CaT wavelength range. 
In the upper panel is the unskysubtracted spectrum and in the lower panel is the 
result of automated sky subtraction. The CaT lines are marked. }
\label{fig:lrexample}
\end{center}
\end{figure*} 

\subsection{Sky subtraction and wavelength calibration}

An example of a spectrum produced by the GIRBLDRS pipeline is shown in
Fig.~\ref{fig:lrexample}.  The numerous skylines visible in this part of the 
spectrum not only serve as an independent check on the overall wavelength 
calibration, but also enable an update of the wavelength calibration of
the individual spectra.

Since the CaT lines only occupy a limited part of the LR spectra we
optimised the sky subtraction and wavelength refinement for the range
8400\AA\,-- 8750\AA\,.

The first step is to combine all the sky spectra (typically 10--20 sky
fibres were allocated per field) using k-sigma clipping to remove
spurious features and obtain an average sky spectrum. The result is
then split into continuum and sky line components using an iterative
k-sigma clipped non-linear filter (a combination of a median and a
boxcar).  The average ``sky-line'' spectrum is then used to define a
sky-line template mask, in order to isolate those regions of the sky
spectrum with significant features and mask out the remainder.

The processing of individual object spectra then proceeds as follows:

i. each spectrum is filtered as above to split the spectra into a line
and continuum component, but this time additionally masking out those
regions affected by sky lines to allow a more accurate definition of
the continuum;

ii. the object line spectrum, which includes sky lines, is then
cross-correlated with the masked line component of the average sky
spectrum. This provides an accurate differential wavelength
update. The object line spectrum is then (re)interpolated to be on the
same wavelength scale as the average sky spectrum;

iii. for the sky subtraction we compare the masked sky-line and
object-line spectra, and find the optimum scale factor and profile
matching kernel that produce the minimum average absolute deviation
(L1 norm). This is applied to the line-only spectra.

The optimum scaling factor, which in this case is chosen to minimise
the L1 norm rather than the commoner L2 norm to reduce sensitivity to
non-Gaussian outliers, is derived using a simple grid search with
progressively finer step size.  As noted previously, a mask is used to
isolate the relevant regions of the sky spectrum to match to a
template.  By first removing the continua from both sky and object,
more emphasis is placed on minimising the impact of sky line
residuals.

iv. finally the object continuum is added back to the
wavelength-updated object-line spectrum, the sky continuum is removed
used the sky-line scaling factor and the sky-subtracted spectrum is
saved for the next stage in the processing. Of course, implicit in the
sky correction is the reasonable assumption that the derived scale
factor for both sky lines and sky continuum is the same.

The sky subtraction process described above involves some key
components which are crucial to achieve good results.  Accurate
wavelength registration is absolutely vital for good sky subtraction
and is facilitated by the presence of copious numbers of strong sky
lines.  These sky lines are unresolved at this resolution and at the
signal-to-noise achieved (see later) which readily enables sub-\kms
precision in wavelength alignment.  This also has the added advantage
of ensuring that systematic offsets due to wavelength calibration for
velocities from different observations are negligible.

As a final step in this process, all the average sky spectra 
from each FLAMES observation, are cross-correlated with a chosen reference 
sky spectrum and used to put all the observations on the same internal system. 
This is done to avoid possible systematic differences between observations 
taken at different times.

The effects of combining the sky spectra to form an average sky and 
re-interpolating the object spectrum to this average wavelength system, almost 
invariably results in a slight mismatch between the spectral line
profiles of the object and average sky spectra. This is circumvented by
applying a Hanning smoothing kernel to each in turn and finding which
combination of smoothed and un-smoothed gives the best results (as
determined by the optimum scale factor).  More sophisticated adaptive
kernel matching \citep[e.g.,][]{alard1998}
is probably unwarranted in this case.

\subsection{Velocity and equivalent width estimation} \label{subsec:velew}

Our goal is to produce a robust automatic procedure that gives
close to optimum results in terms of signal:to:noise and also produces
minimal systematic bias as a function of EW.  

The first stage of the process is to estimate the continuum in a
similar manner to that described in the previous section.  Each
spectrum is split into a line-only, and smoothly varying
continuum-only component, using an iterative k-sigma clipped
non-linear filter.  This time the region around each CaT line is
masked out, prior to filtering, to prevent the continuum tracking the
wings of strong CaT lines.  The effective scale length of the
filtering is set to $\sim$ 15\AA\ which, in conjunction with the
masking and iterative clipping, is sufficient to follow continuum
trends without being affected by the presence of strong lines.  At the
same time an estimate of the overall signal-to-noise in the continuum
is made by measuring the median continuum level in the CaT region and
the pixel-to-pixel noise covariance matrix in regions containing no
lines.  The latter is needed to correct the apparent random noise for
the cumulative smoothing effects of spectral interpolation and
resampling, which we find typically results in a factor of $\approx$2
overall random noise reduction.

\begin{figure*}
\begin{center}
\includegraphics[width=120mm,angle=270]{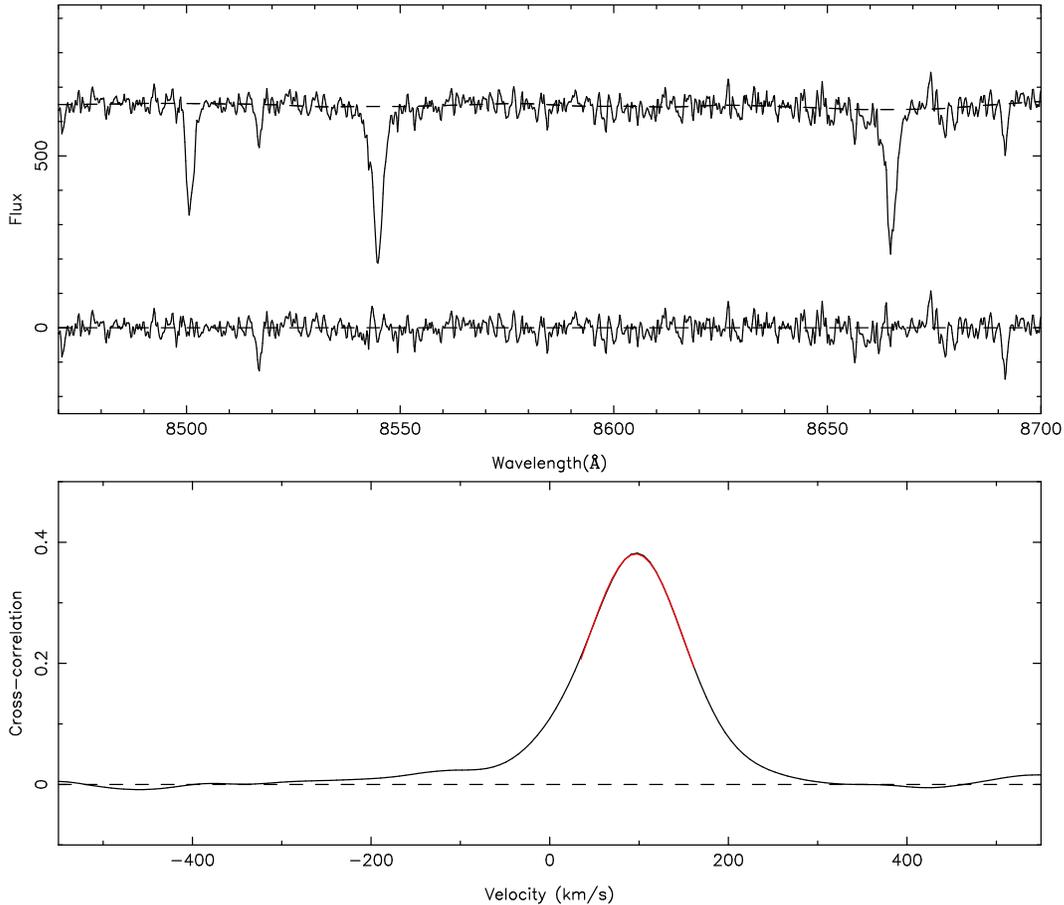}
\caption{Top: the upper spectrum shows the CaT region for a K-giant in
the Sculptor dSph (V$=$17.7, S/N $\approx$ 40/\AA, 
derived [Fe/H]$_{\rm CaT} = -$1.50 dex), 
showing an example of automated continuum fitting (dashed line); the lower 
spectrum shows the residuals after continuum removal and subtracting 
Gaussian model fits to the three CaT lines. Bottom: the derived
cross-correlation function and associated Gaussian fit around the peak region ($\pm$ 50 \kms about the peak).
The fit is so good that it is indistinguishable from the observations.
}
\label{fig:cross}
\end{center}
\end{figure*}

After normalising by the computed continuum the velocity is estimated
by cross-correlating each spectrum with a template.  This template is
constructed from a zero-level continuum superimposed on three Gaussian
absorption lines located at the vacuum rest wavelengths of the CaT
lines.
Most of the weight in the least-squares fit comes from the core of the line 
which is sufficiently Gaussian-like in LR data, that a Gaussian fit 
provides an estimate with effectively a minimal rms error.  Using more 
complex line profiles with correspondingly greater numbers of free 
parameters generally makes the rms error worse and is also more prone to 
wildly unstable solutions due to the inevitable presence of occasional
artefacts in the data.
The Gaussian line depths are scaled in the ratio 3:5:4 to
reflect the true relative strengths of CaT lines and all are set to
have a full width at half maximum (FWHM) $=$ 2.35\AA\,.  An example of
a continuum fitting to a Sculptor dSph K-giant spectrum together with
the computed cross-correlation function and Gaussian fit to the peak
is shown in Fig.~\ref{fig:cross}.

An accurate estimate of the position of the cross-correlation peak is
made by fitting a Gaussian to a localised region around the peak.
This velocity is then used to define the wavelength region around each
CaT line to use for EW estimation. We estimate the EW in two ways. The
first consists in simply summing the flux contained in a region
centred on each CaT line.  After some trial and error we settled on a
region 15\AA\ wide centred on each line as a reasonable trade-off
between including all the line flux and minimising the noise. To
derive the second estimate we fit individual unconstrained Gaussian
functions to each CaT line over the same wavelength region (see
Fig.~\ref{fig:cross}).  This also allows a semi-independent check on
the accuracy of the derived velocity by providing three separate
velocity measures with associated errors.  The weighted sum of these
velocity errors provides the basic error estimate for the velocity
derived from the cross-correlation.  We prefer to use the latter
method for the velocity estimate since it is effectively a constrained
model fit.  As a final step the derived velocity is corrected to the
heliocentric system.

\begin{figure*}
\begin{center}
\includegraphics[width=100mm,angle=270]{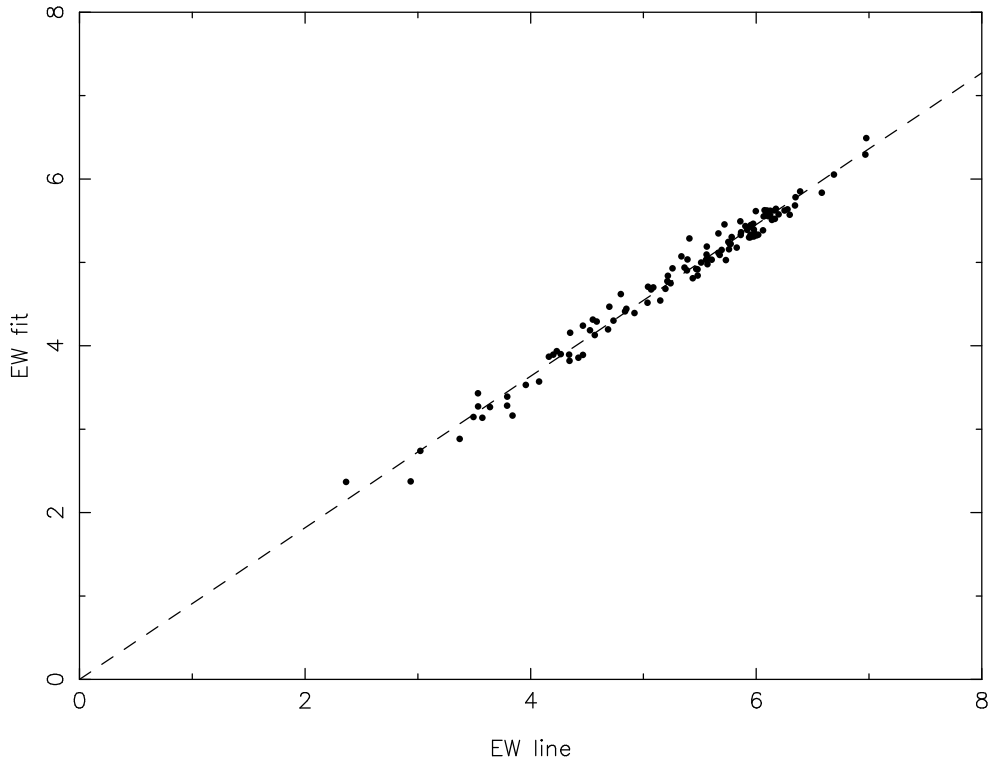}
\caption{A comparison of the EW obtained from profile fitting (EW fit)
versus the EW from the direct integral of the lines (EW line), as 
an example of the 
relationship between the two}.
\label{fig:ewcmp}
\end{center}
\end{figure*}

The combined EWs for CaT lines \#2 and \#3 ($\lambda_{\rm 8542}$,
$\lambda_{\rm 8662}$) for both the integral and Gaussian fits
are then compared and used to compute an overall correction to the
Gaussian fit.  This is necessary since the real line profile is a
complex function of many parameters, and in particular, the dampening
wings visible in strong lines are distinctly non-Gaussian in
appearance. This means 
that the observed CaT lines have non-Gaussian wings 
which are progressively more visible as the EW increases. To compensate 
for this we compare the ensemble Gaussian fit to the conventional EW 
integration for each dataset analysed, by finding the best fit slope that 
links the two sets. 
This correction is equivalent to computing the average
overlap integral between the real line profiles and the Gaussian fits
and is accomplished by measuring the gradient between the two as a
function of EW. 
By using all the data to do this we 
introduce no additional rms error and remove the majority of the systematic 
bias as a function of EW.  We also automatically take care of any generic 
spectrograph-induced line profile abnormalities.  For the range of
EWs considered here this method works reliably (see Fig.~\ref{fig:ewcmp}).
In deriving this correction we neglect line \#1
($\lambda_{\rm 8498}$) as it is the weakest and would only add
noise to the determination.

In the rest of our analysis we use the Gaussian-derived EW estimator
since although the simple integral is unbiased it is also
significantly noisier than a Gaussian fit.  We show in the next
section how the measurements of the continuum level and the random
noise error in the continuum can be used to estimate errors in the EWs
and velocities.

\subsection{Error bounds for velocities and equivalent widths}

For detailed abundance work most lines of interest are weak \eg, EW
$\approx$ 100m\AA.  Notable exceptions are the CaT lines which are
generally heavily saturated and on the damped part of the
curve-of-growth.  Turbulence plus rotation of late-type giants
typically only broaden the line profiles by $\approx$ a few \kms,
hence the profile of the CaT lines, which are completely unresolved at
R$=$6500, are dominated by intrinsic broadening due to saturation
(typically FWHM 2\AA\ -- 3\AA\,) and to a lesser extent by the
resolution of the spectrograph (FWHM $\approx$1.3\AA\,).

Despite the dampening wings, to first order the LR CaT lines can be
reasonably well approximated by Gaussian functions (\eg see
Fig.~\ref{fig:cross}) and we can use this to gain some insight into
the limiting factors that determine the accuracy of the velocity and
EW measurements.

Here we define the resolution as FWHM $= 2.35 \sigma$ where $\sigma$
is the Gaussian profile equivalent scale parameter.  Since the total
line flux is then
 $I_p \sqrt{2 \pi} \sigma = I_p \times$ FWHM $\times 1.07$, where
 $I_p$ is
the ``peak'' flux, this implies that line saturation (\ie $I_p = C$,
where $C$ is the continuum level/\AA\,) occurs when EW $\approx$ FWHM.
The intrinsic FWHM of weak lines in these late-type giants is
typically only a few \kms, \ie lines with EWs of $\approx$100m\AA\ and
above are saturated.  As noted previously, the CaT lines are heavily
saturated and typically have EWs well above 1\AA\,.

To a reasonable approximation, the noise in the continuum ($\sigma_n$
per \AA\,) due to sky plus object dominates, and over any individual
line $C$ can be taken to be constant over the lines of interest.
Therefore, the EW and its error due to random noise, $\Delta EW$, are
given by
\begin{equation}
 EW \ = \ {\eta \over C}  \ \ \ \ \ \ \ \ 
   \Delta EW \ = \ {\sigma_n \sqrt{w} \over C} = {\sqrt{w} \over S/N }
\end{equation}
where the total line flux is $\eta$, $S/N$ is the signal-to-noise per
\AA \ and $w$ is the effective width (\AA) the line is integrated
over.

For a fitted Gaussian profile 
\begin{equation}
 w = \sqrt{4\pi} \sigma \approx 1.5 \ {\it FWHM}
\end{equation}
hence 
\begin{equation}
\label{eq:deltaew}
 \Delta EW \ = \ {\sqrt{1.5 \ {\it FWHM}} \over S/N} 
\end{equation}
The FWHM of the line and continuum signal-to-noise are the primary 
abundance error drivers from a random noise point-of-view.  For example,
the two strongest CaT lines, $\lambda_{8542}, \lambda_{8662}$, used 
in our analysis have FWHM at a resolution of R$=$6500 of 2\AA\ --3\AA\,,
which for a continuum signal-to-noise of 10/\AA\, implies a lower bound
on the combined EW error of $\approx$0.3\AA\,.

\vspace{5mm}

In a similar way we can place constraints on the accuracy of measuring
velocities.  For Gaussian-like line profiles, which are good approximations 
even for saturated lines like the CaT, the minimum
variance bound on the error in the estimated line position $\hat \lambda$
is given by
\begin{equation}
var\{\hat \lambda\} \ = \ {\sigma^2 \over \eta} \ {\sigma_n^2 \sqrt{16 \pi} 
                           \ \sigma \over \eta }
\end{equation}
where the line flux is $\eta$ and where the noise in the continuum, 
$\sigma_n$ (/$\AA$), dominates and can be taken to be a constant over the 
region 
of interest \citep[see][for more details]{irwin1985}.  
Rewriting this in terms of the FWHM of the line, the EW 
and the continuum signal-to-noise, leads to
\begin{equation}
\Delta \hat \lambda \ \approx \ {{\it \ FWHM}^{3 \over 2} \over EW \ S/N}
\end{equation}
where all measurements are per \AA.

As an example, for observations at R$=$6500, the FWHM of the strongest CaT 
line is typically between 2--3\AA\,, while the EW of 
these lines is typically $\approx$2\AA\,,implying accuracies of wavelength
centring at a continuum signal-to-noise of 10/\AA\,of around 0.2\AA per
line, or equivalently $\approx$5 km/s using all three CaT lines.
This is the minimum signal-to-noise we consider acceptable in our analysis
and is achievable on VLT FLAMES for V$=$20 objects in 3600s of integration.

\begin{figure*}
\centering
\begin{tabular}{cc}
\includegraphics[width=60mm]{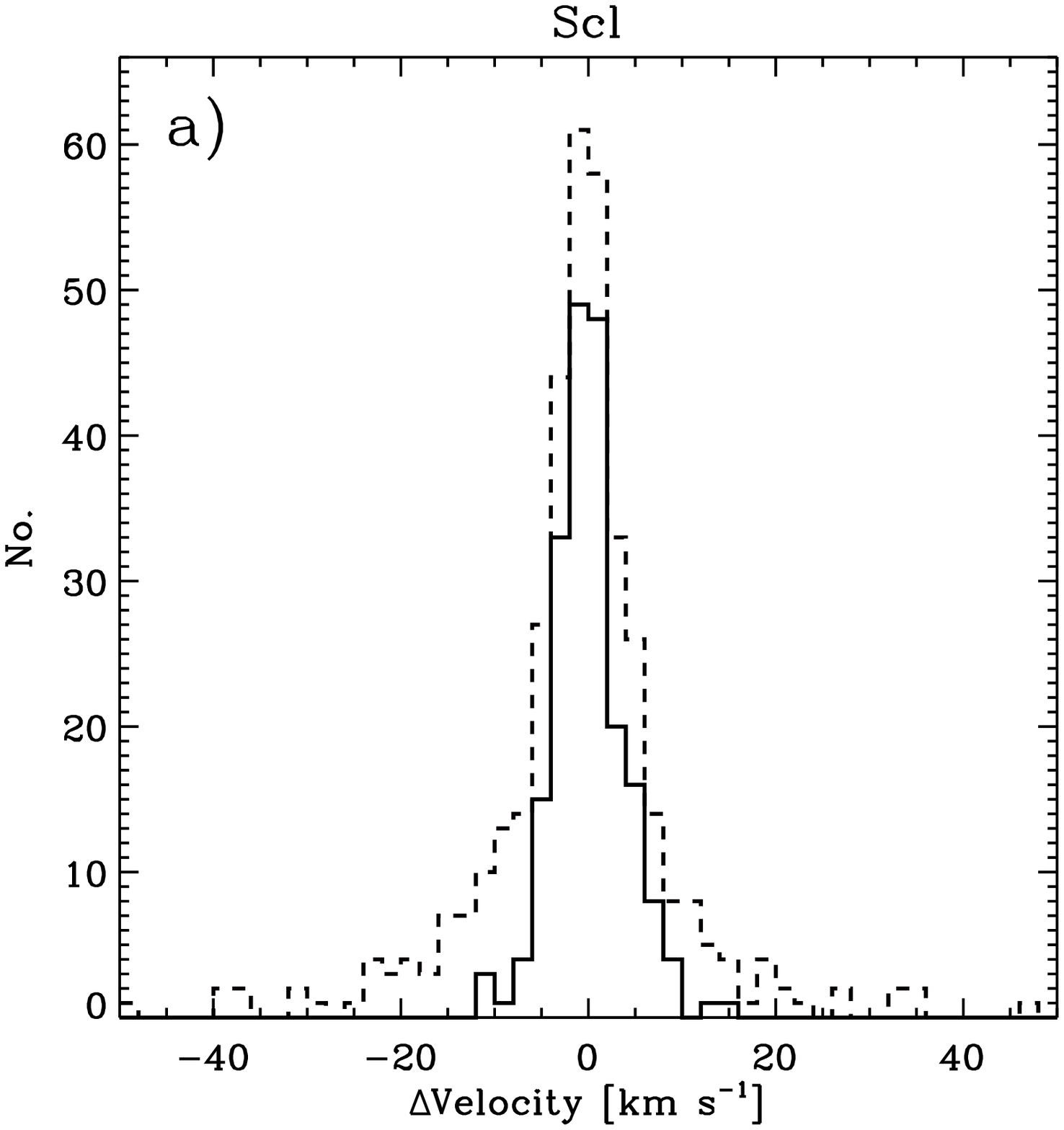}&
\includegraphics[width=60mm]{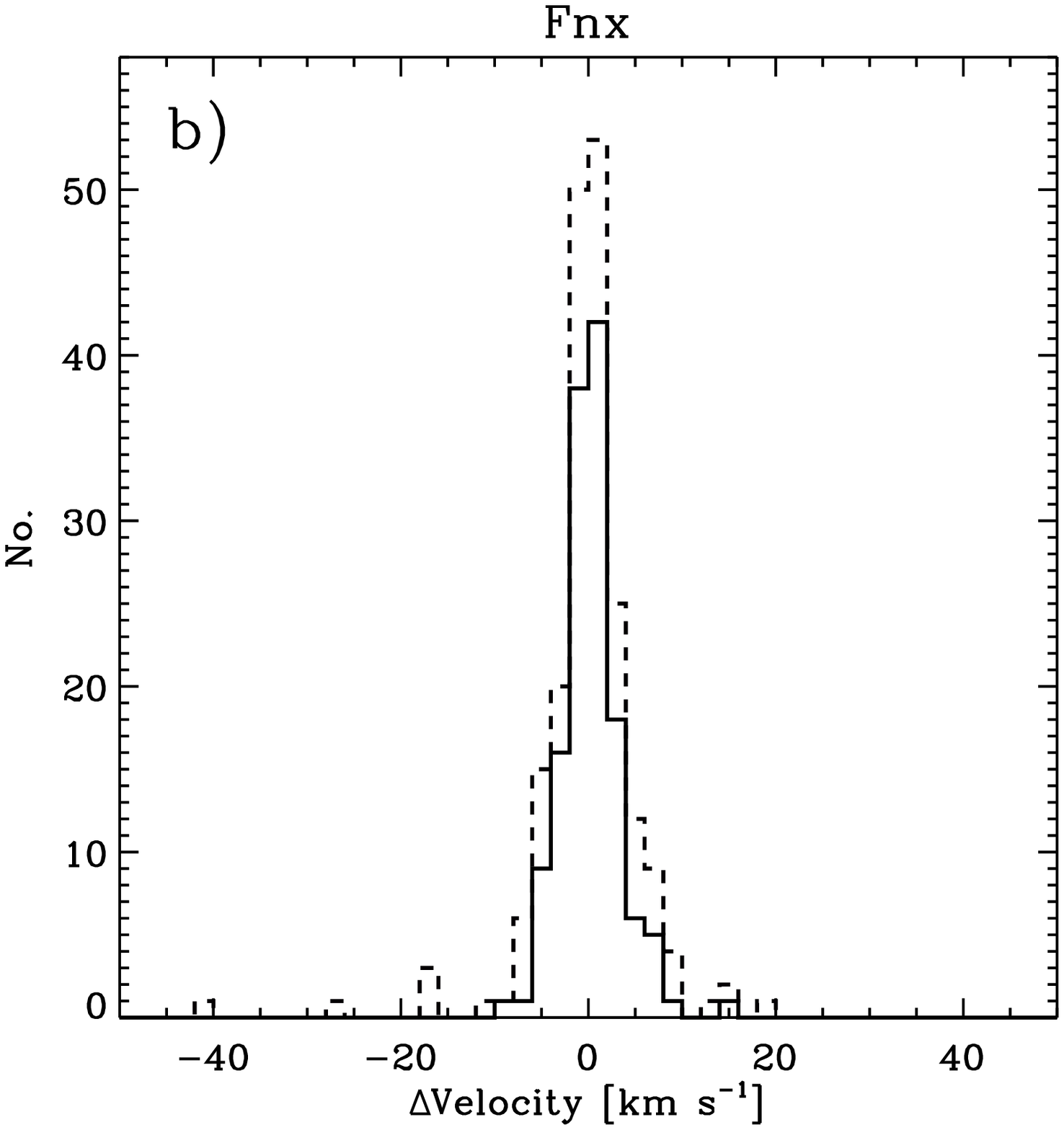}\\
\includegraphics[width=60mm]{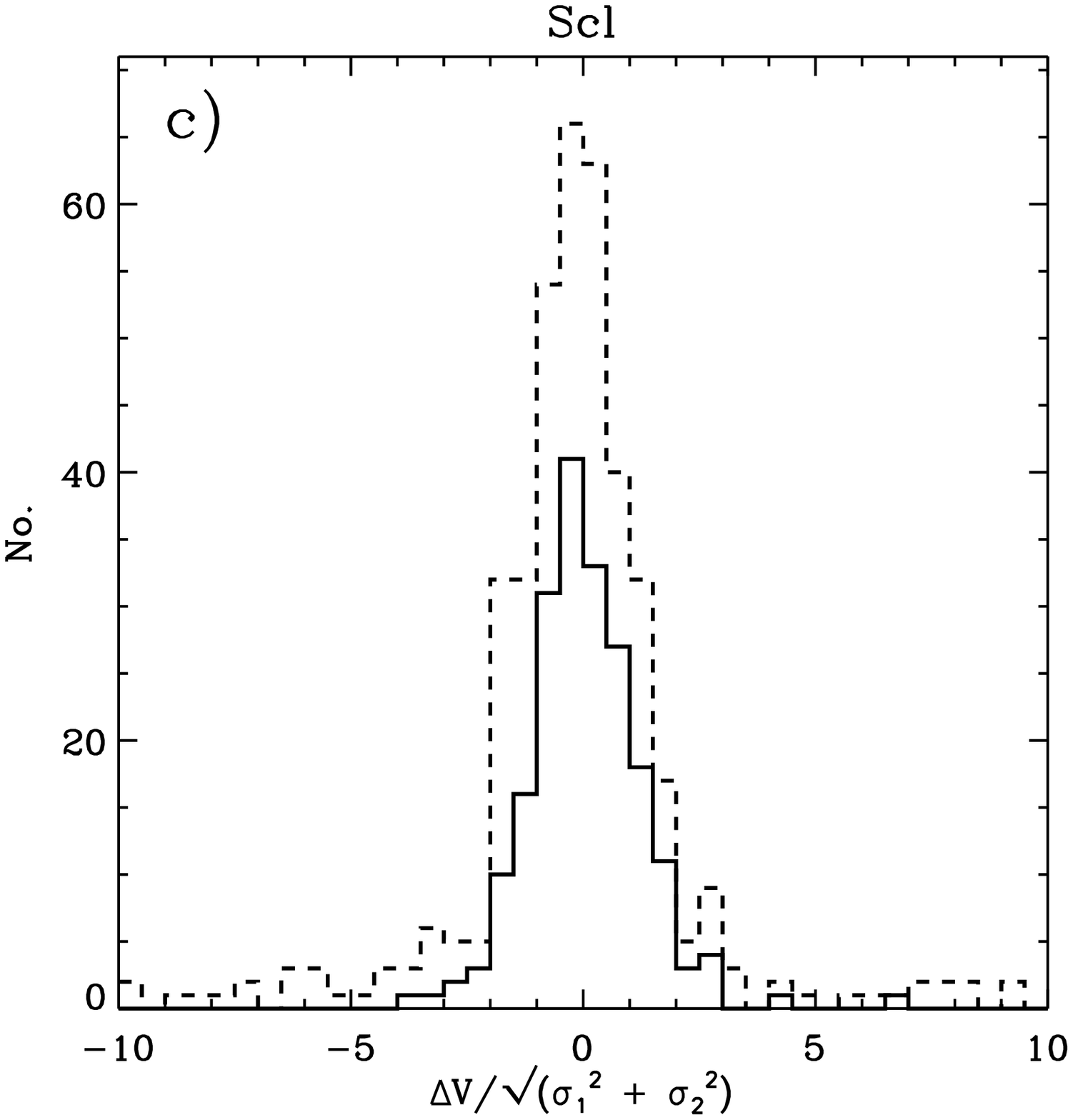}&
\includegraphics[width=60mm]{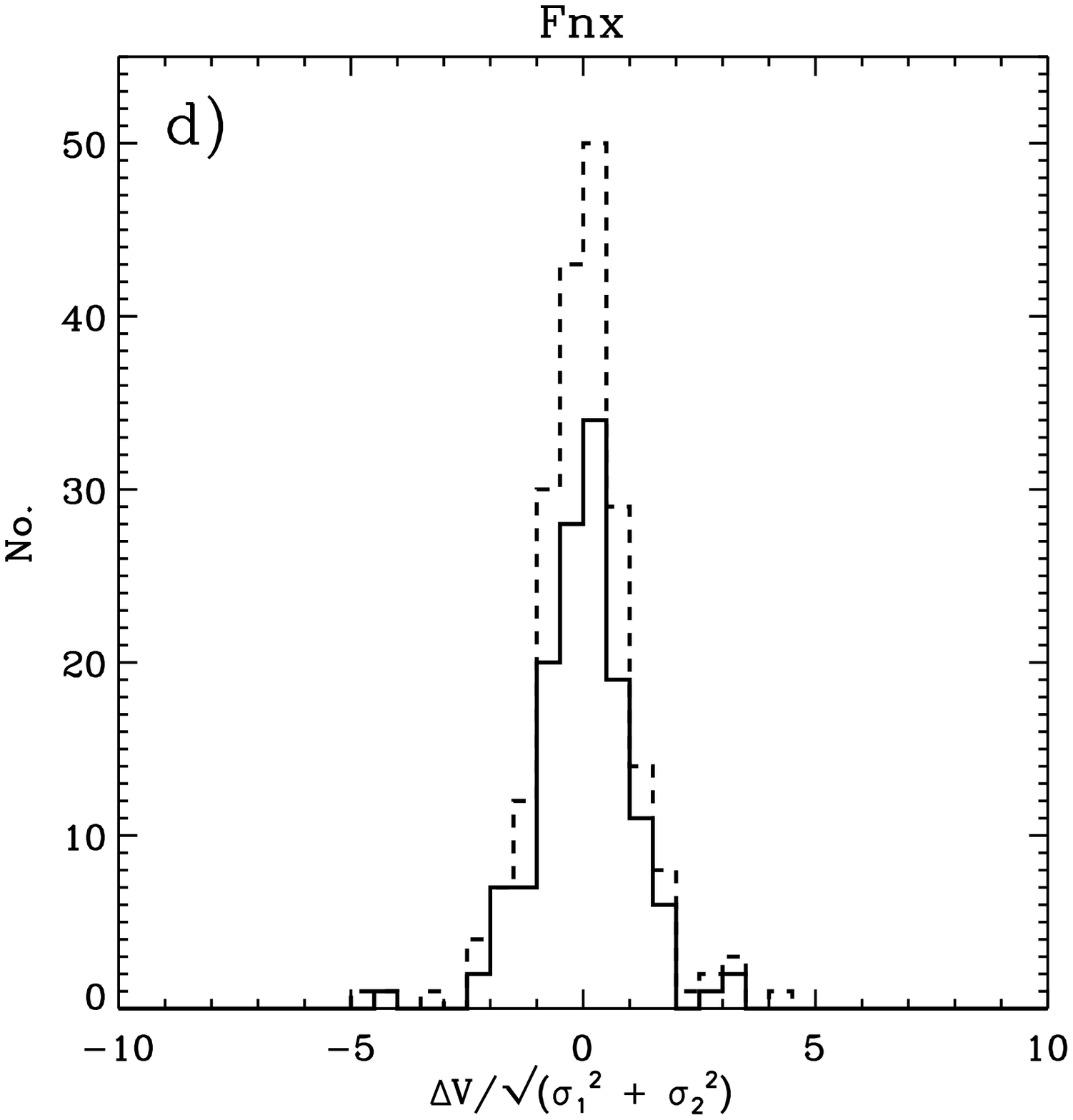}\\
\includegraphics[width=60mm]{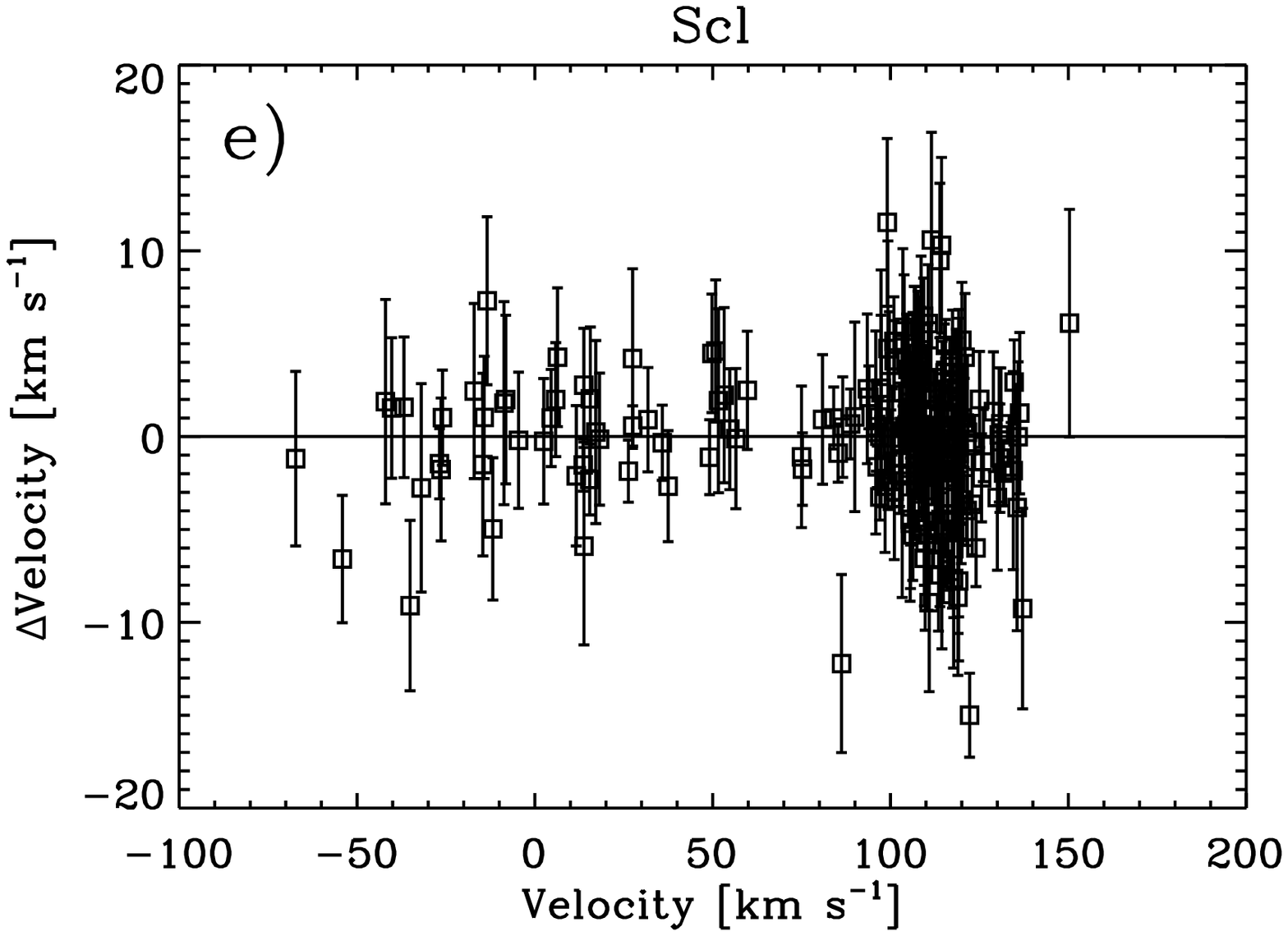}&
\includegraphics[width=60mm]{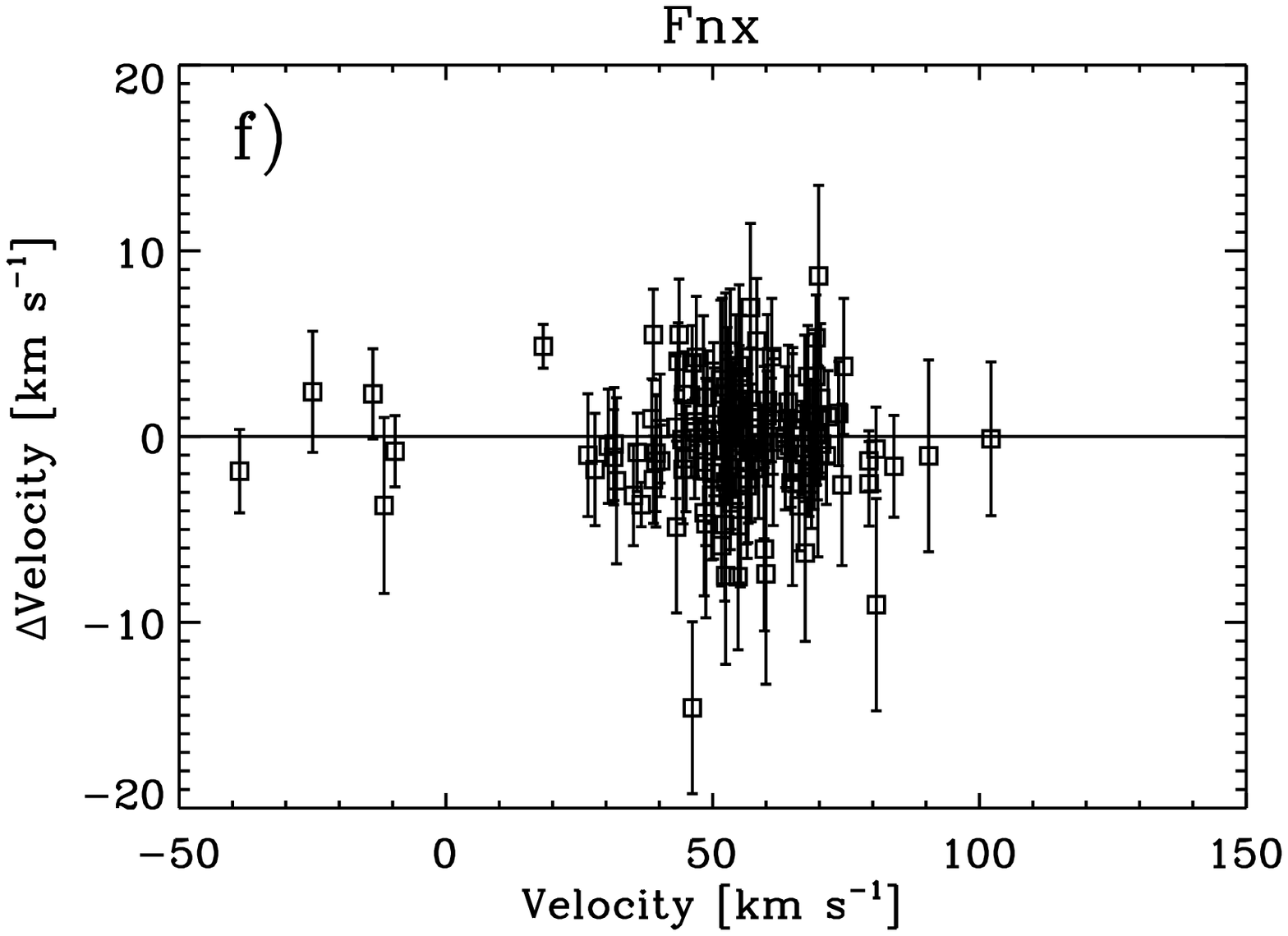}
\end{tabular}
\caption{Comparison between velocity measurements for stars with double measurements in the 
Sculptor (left) and Fornax (right) dSphs. Panels a, b) Distribution 
of velocity differences for all the stars (dashed line, 428 stars for Sculptor and 209 for Fornax), 
and for the stars with S/N per \AA\,$>$ 10 and 
estimated error in velocity $<$ 5 \kms for each measurement (solid line, 203 stars for Sculptor and 138 for Fornax). 
The weighted mean velocity, 
$rms$ dispersion and scaled median absolute deviation from the median (1.48*MAD $\equiv$ a robust $rms$ \eg 
Hoaglin et al. \citeyear{hoaglin1983})
 are: $-0.5 \pm 1.3$ \kms, $25.1 \pm 1.0$ \kms, $6.7 \pm 0.2$ \kms (dashed line) and $0.1 \pm 0.3$ \kms, $3.1\pm 0.4$ \kms, 
$3.1 \pm 0.4$ \kms (solid line) for Sculptor; $1.4 \pm 2.0$ \kms, $27.7 \pm 1.4$ \kms, $3.2 \pm 0.2$ \kms (dashed line) 
and $-0.01 \pm 0.36$ \kms, $2.6\pm 0.4$ \kms, $2.4 \pm 0.4$ \kms (solid line) for Fornax. 
Panels c, d) As above but now the velocity difference is normalised by the predicted error.  With these S/N and velocity error cuts 
the measured error in the velocity distribution is very close to the expected unit variance Gaussian
(standard deviation = 1.2 and MAD = 1.1 for Sculptor; standard deviation = 1.0 and MAD = 0.9 for Fornax). 
Panels e, f) Comparison of velocities for stars with S/N per \AA\,$>$ 10 and error in velocity $<$ 5 \kms.}
\label{fig:vel_cc}
\end{figure*} 

\begin{figure*}
\centering
\begin{tabular}{cc}
\includegraphics[width=70mm]{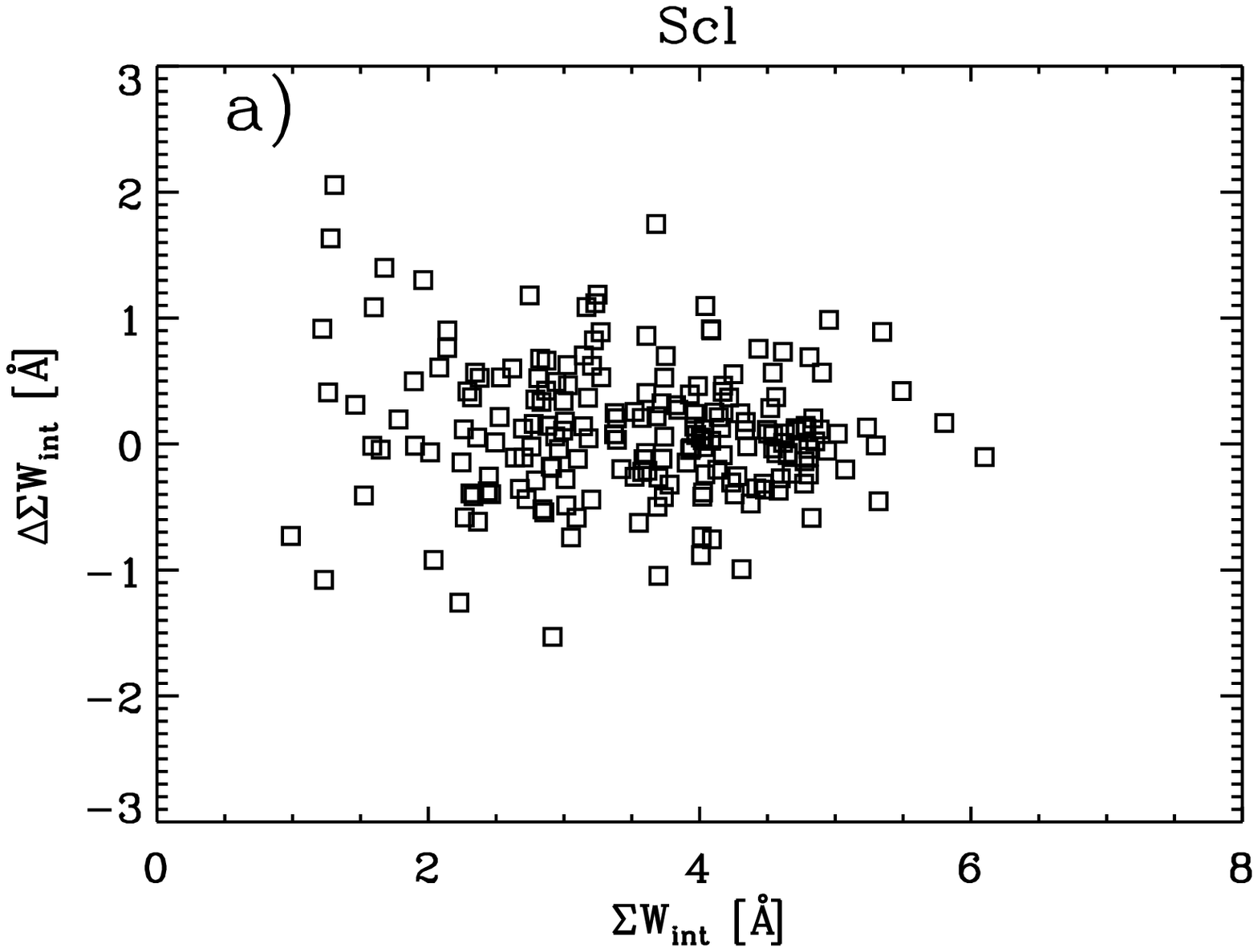}&
\includegraphics[width=70mm]{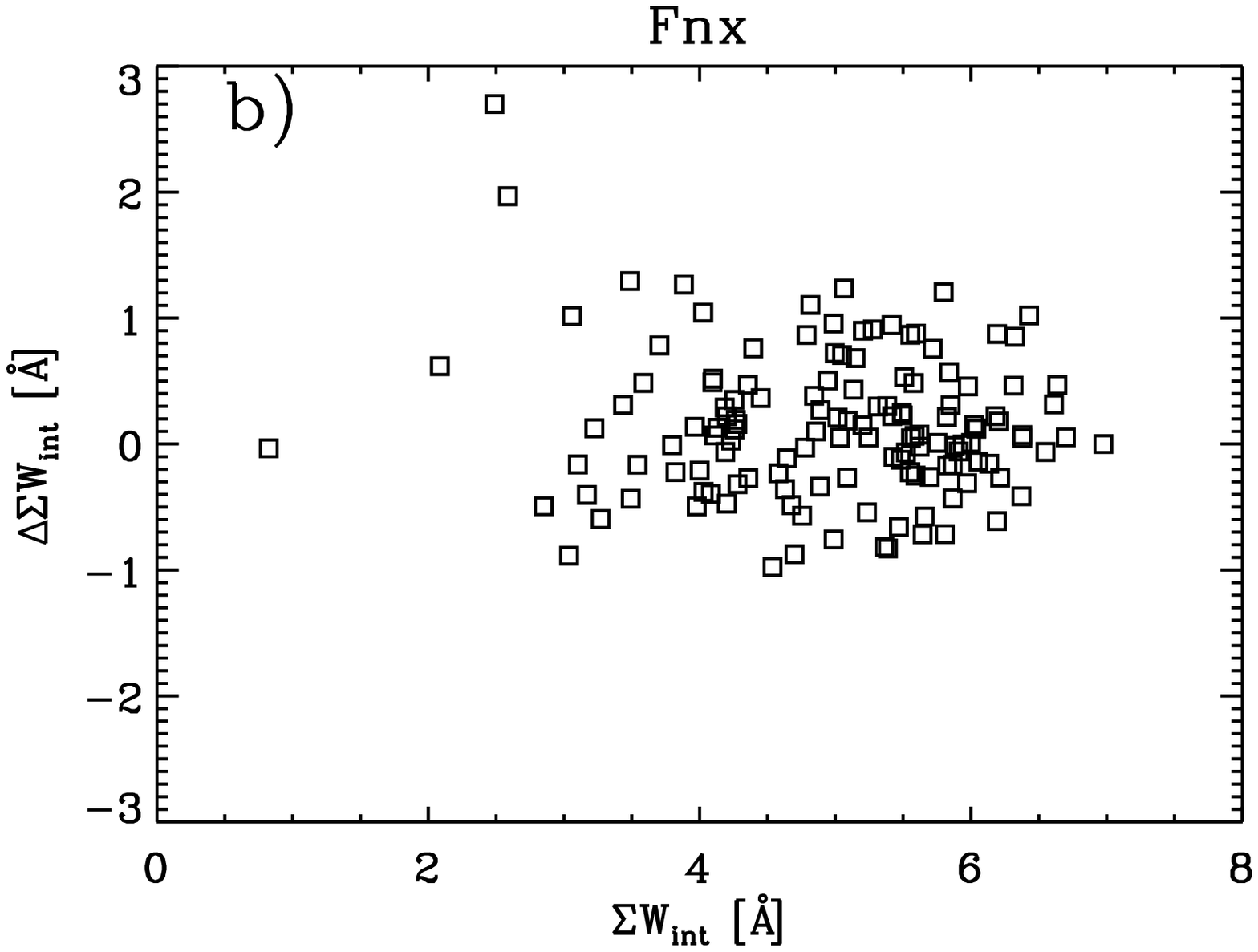}\\
\includegraphics[width=70mm]{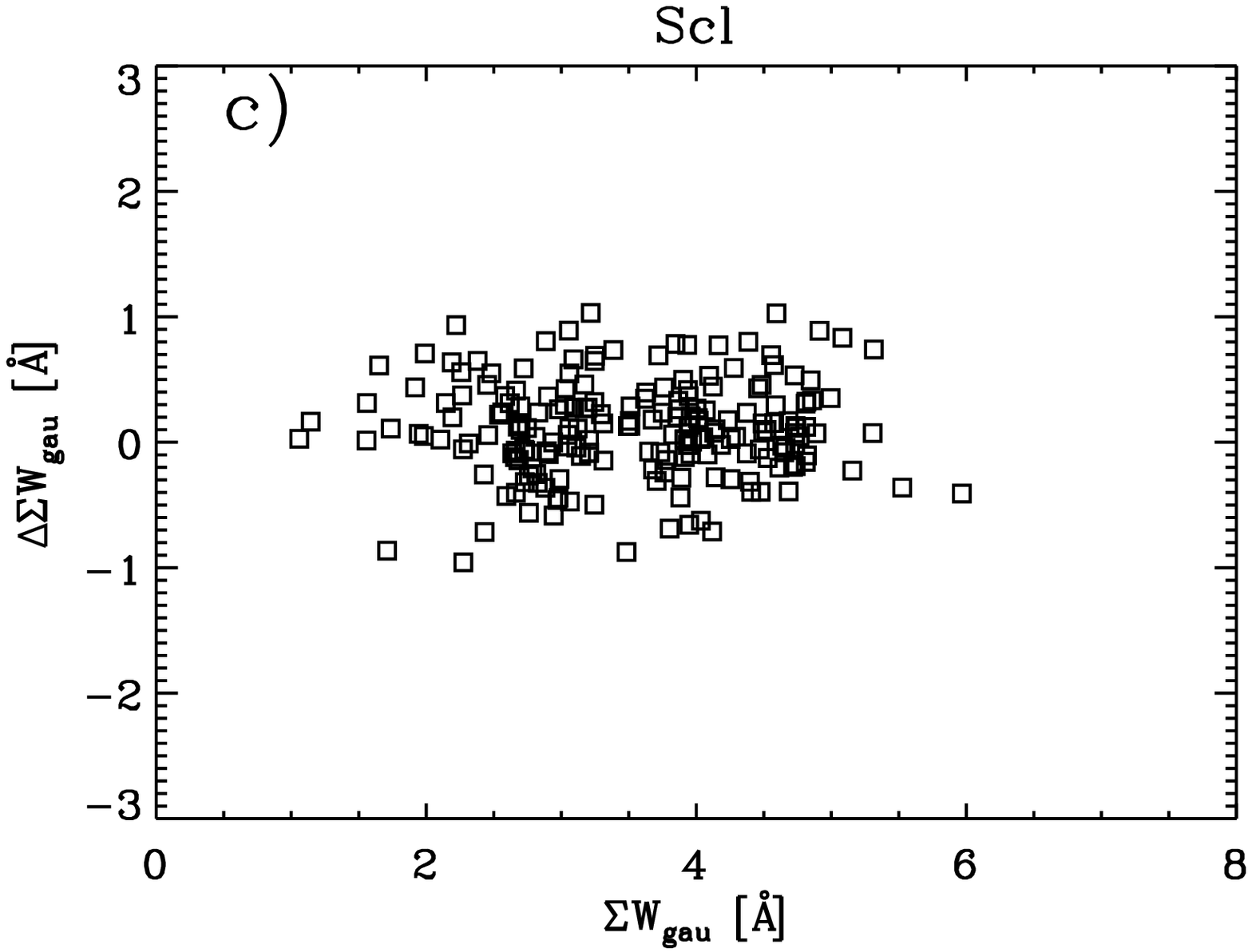}&
\includegraphics[width=70mm]{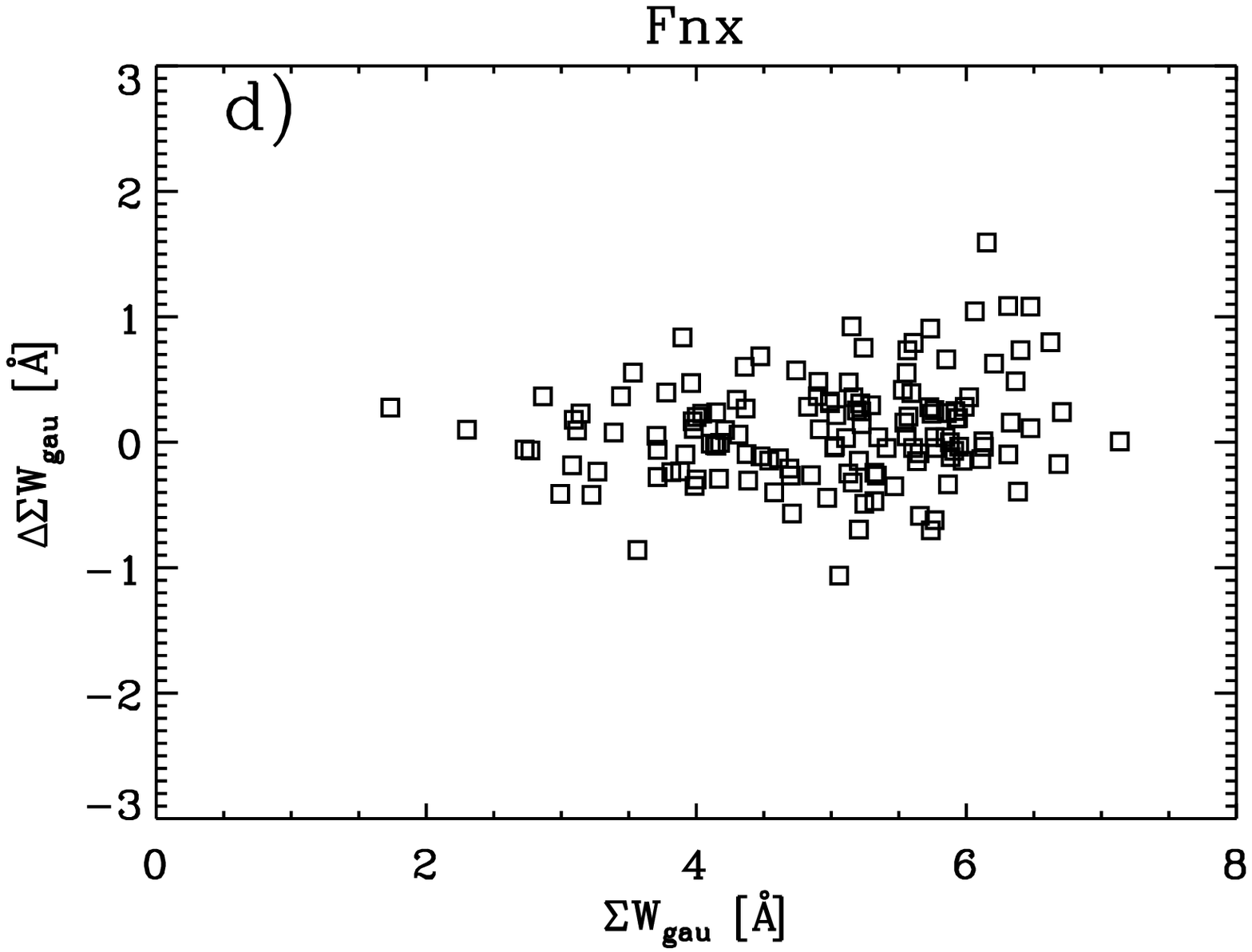}\\
\includegraphics[width=70mm]{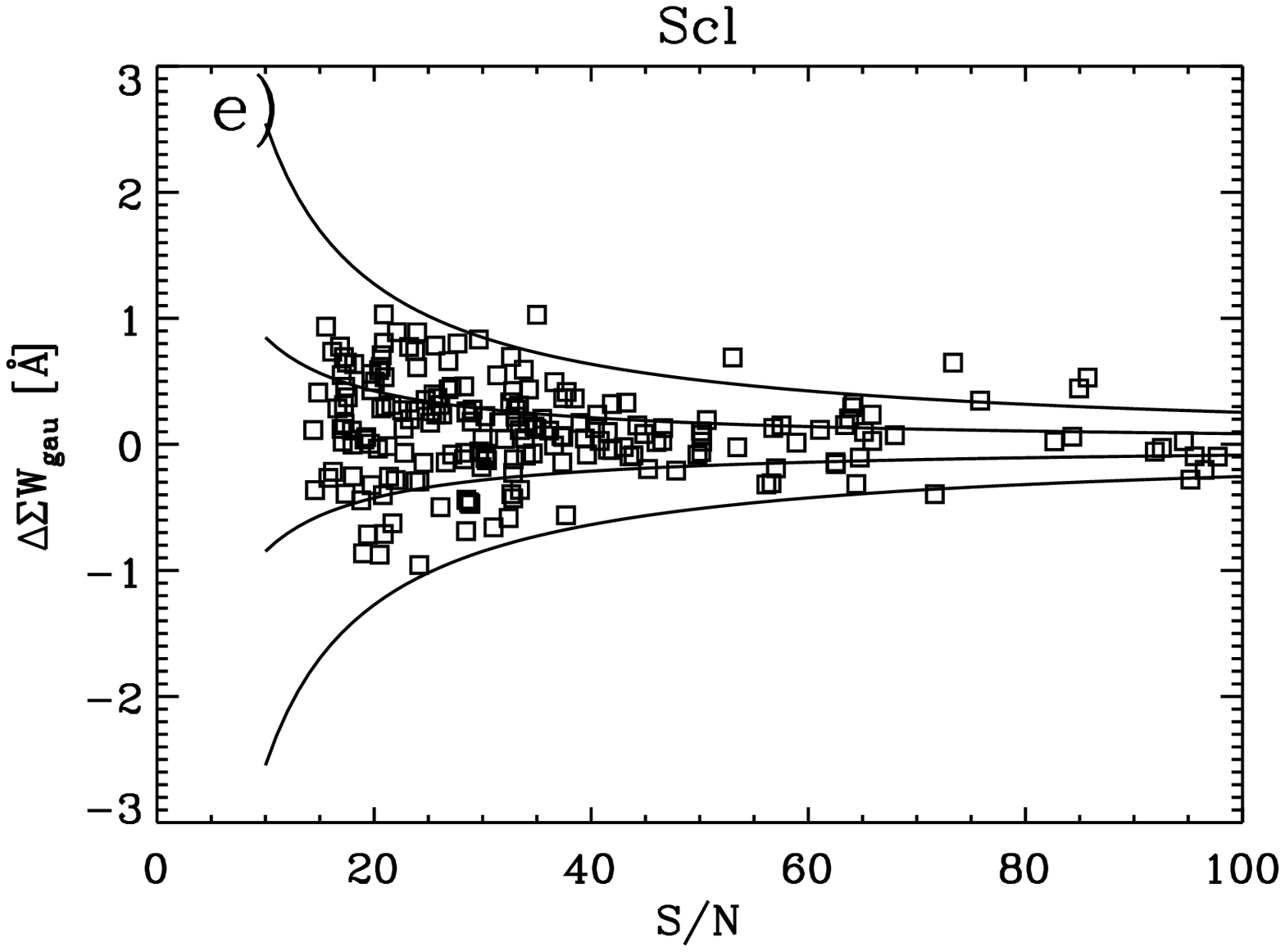}&
\includegraphics[width=70mm]{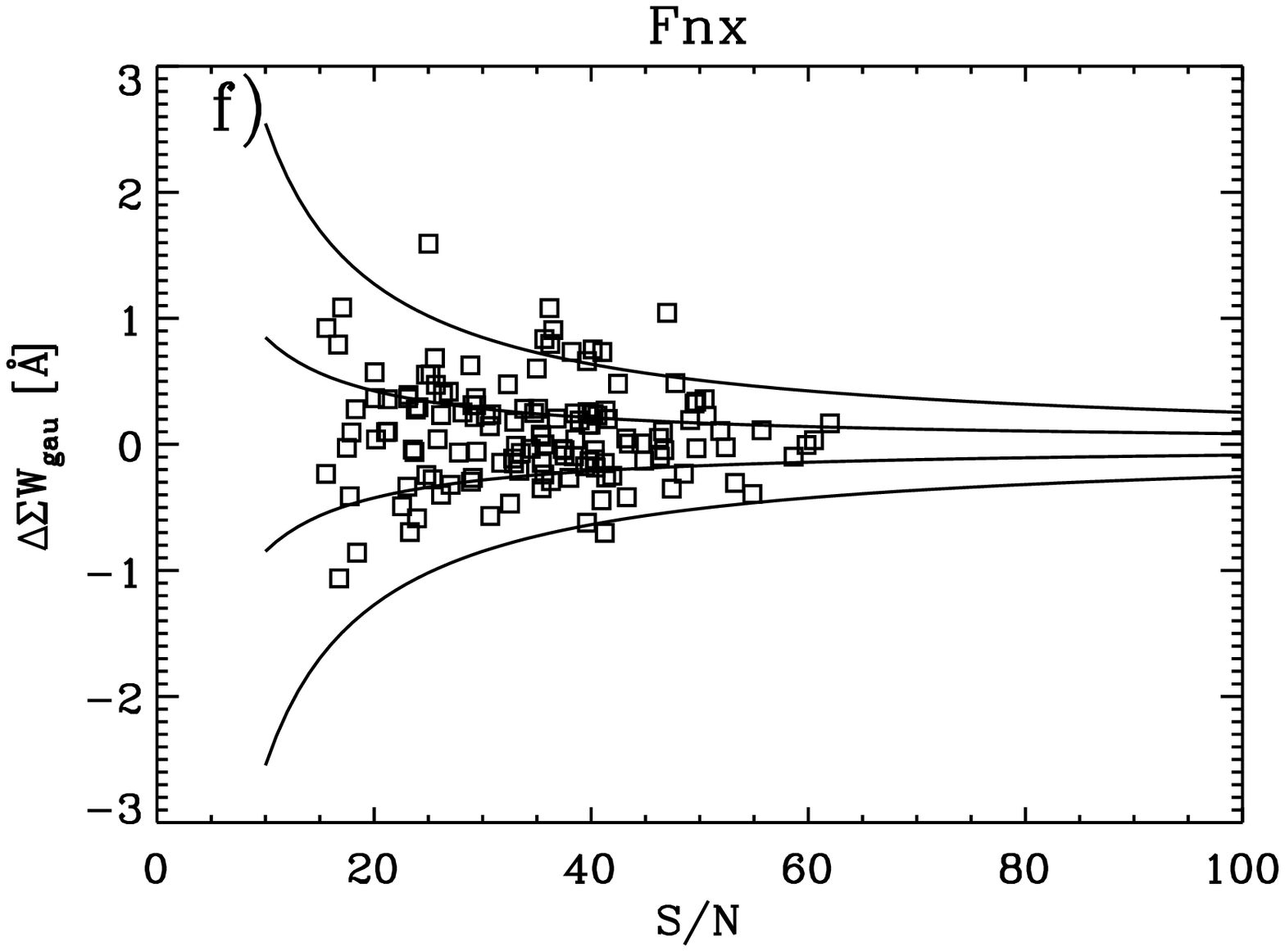}
\end{tabular}
\caption{Comparison between summed EW (EW$_2$+EW$_3$) measurements of 
CaT lines from integrated flux (a,b) and Gaussian fitting (c, d, e, f) for stars with double measurements, 
S/N$>10$ and error in velocity $<$ 5 \kms  in the Sculptor (left, 203 stars) and Fornax (right, 138 stars) dSphs. 
 The solid lines in panel (e, f) indicate the 1 and 3 $\sigma$ 
region for an error in summed EW given by $\sigma_{\Sigma W} = 6 / (S/N)$ (see text). 
Assuming this error, the weighted average, $r.m.s.$ and MAD for the differences in $\Sigma W$ from 
Gaussian-derived estimator (panels c, d) is $0.04 \pm 0.04$ \AA\,, $0.30 \pm 0.05$ \AA\,, $0.32 \pm 0.05$ \AA\, 
for Sculptor and $0.07 \pm 0.05$ \AA\,, $0.36 \pm 0.05$ \AA\,, $0.34 \pm 0.05$ \AA\, for Fornax.
}
\label{fig:ew_comp}
\end{figure*} 

\subsection{Errors for velocity and equivalent width from repeated measurements}
The number of
independent measurements for each dataset is 1740 for Sculptor and
1359 in Fornax.  
For Sculptor we have 464 stars observed only once, 428 stars observed
twice, 73 stars observed 3 times, 39 observed 4 times and 9 observed 5
times.  For Fornax we have 816 stars observed only once, 209 stars
with double, 23 stars with triple, 14 stars with quadruple
measurements.

We test for the reliability of our velocity and related errors by
analysing the distribution of velocity differences from double
measurements.  The $j$th observed velocity $v_{ji}$ for a star $i$ can
be considered a random variable which follows a
 Gaussian distribution centred around the
true value $v_{\rm true,\it i}$ with a dispersion given by the
velocity error $\sigma_{ji}$.  The difference between two repeated
measurements $v_{1,i}$ and $v_{2,i}$, $\Delta v_i= v_{1,i}-v_{2,i}$,
is a random variable following a Gaussian distribution centred around
zero and with dispersion given by $\sigma_i = \sqrt{ \sigma_{1,i}^2 +
\sigma_{2,i}^2}$.  Thus, if both velocities and their errors are
correctly determined,
 the distribution of velocity differences $\Delta v_i$ normalised by
 $\sigma_i$ should be a
Gaussian with mean zero and dispersion unity. Figure~\ref{fig:vel_cc} 
shows that if we take into consideration all
the stars with repeated measurements both for Sculptor and Fornax the
distribution has large tails of stars with $\Delta v/\sigma > 3$. If
we restrict ourselves to the stars with S/N per \AA\,$>$ 10 and error
in velocity $<$ 5 \kms, then the resulting distribution is very close
to a Gaussian. For Sculptor the resulting standard deviation is 1.2
and the scaled Median Absolute Deviation (MAD) is 1.1, with just 5
stars with $\Delta v/\sigma > 3$ (2\%).  For Fornax the standard
deviation is 1.0 and the scaled MAD is 0.9.

Hereafter we consider a S/N per \AA\,$>$ 10 and an error in velocity
$<$ 5 \kms as the minimum requirements for a reliable determination of
velocity and EW.

Figure~\ref{fig:ew_comp}a,b,c,d shows
the comparison of summed EW for the two strongest CaT lines, $\Sigma W
= EW_2 + EW_3$, which we use as indicator of metallicity (see Sect.~\ref{sec:calibration_gcs}),
for the stars with double measurements in Sculptor and Fornax
respectively.  As mentioned in Sect.~\ref{subsec:velew} the Gaussian-derived estimator
is considerably less noisy than the integral estimator and thus we use
this estimator in all our analysis. As an extra criteria to guarantee
the reliability of the estimated EW measurements we also impose a
cut-off in the difference between the Gaussian-derived and the
integral estimators of less than 2 \AA\,.

We derive the error in $\Sigma W$ for single observations and how this
varies with S/N from the comparison of $\Sigma W$ for stars with
double observations using Eq.~(\ref{eq:deltaew}). Figure~\ref{fig:ew_comp}e,f 
shows that, as expected, the comparison of
$\Sigma W$ improves for increasing S/N. We find that the random error
in $\Sigma W$ from repeated measurements is well represented by
$\sigma_{\Sigma W} \approx 6/(S/N)$, and hereafter we will adopt this
formula as estimate of the errors in the single measurements. This is
a factor $\sim 2$ larger than we estimated from theoretical
calculations; however a larger error is not surprising given the
effect of all the steps involved in the data reduction (e.g. sky
subtraction and continuum estimation). Such an error in $ \Sigma W$
results in a [Fe/H] error of $\sim$ 0.1 dex at a S/N per \AA\,of 20
(see Sect.~\ref{sec:calibration_gcs}).

Finally, the measurements for stars with multiple observations were
combined, weighting them by their error, and this results in 1013
distinct targets for Sculptor and 1063 for Fornax. The final sample
was carefully checked to weed out any spurious objects (e.g., broken
fibres, background galaxies, foreground stars, etc).  Excluding the
spurious objects, and the objects that did not meet our S/N and
velocity error criteria, our final sample of acceptable measurements
consists of 648 stars in Sculptor and 944 in Fornax. Among these, 2
stars in Sculptor and 1 in Fornax did not meet our EW criteria.

As an indication of the good quality of our data, the median S/N and
median error in velocity are 32.1 and 1.6 \kms, respectively, for the
Sculptor dataset, and 24.3 and 1.7 \kms for the Fornax dataset.

\section{The standard CaT calibration with VLT/FLAMES using globular clusters} \label{sec:calibration_gcs}
The next step is to transform the CaT EW into metallicity, [Fe/H].
The dependence of CaT line strength on metallicity is theoretically
difficult to understand, however it has been empirically proved by
extensive calibration using RGB stars in globular clusters
\citep{arm1988, ol1991, arm1991}.  \citet{rutledge1997a} presented the
largest compilation of CaT EW measurements for individual RGB stars in
globular clusters, which \citet{rutledge1997b} calibrated with HR
metallicities, proving the CaT method to be reliable and accurate in
the range $-2.1 \la $ [Fe/H] $ \la -0.6$.

As summarised in \citet{rutledge1997b} the line strength index $\Sigma
W$, which is a linear combination of the EW of individual CaT lines,
depends on [Fe/H], the gravity log$\, g$ and $T_{\rm eff}$.  As log$\,
g$ and $T_{\rm eff}$ decrease going up the RGB, it is possible to
remove the effect of the gravity and temperature taking into account
the position of the star on the RGB with respect to the Horizontal Branch (HB).

\citet{arm1991}, using globular clusters, showed that this is most
efficiently achieved when defining a ``reduced equivalent width'':
\begin{equation}
W' = \Sigma W + \beta \times (V - V_{\rm HB})
\end{equation}
where $V_{\rm HB}$ is the mean magnitude of the HB.
The advantage of using $V-V_{\rm HB}$ over for instance the absolute
$I$ magnitude or the $V-I$ colour is that the slope $\beta$ is
constant, since it does not vary with [Fe/H].  Using $V_{\rm HB}$ also
removes any strong dependence on the distance and/or reddening. The
``reduced equivalent width'' $W'$ is thus the CaT line strength index
at the level of the HB.  With the above definition,
\citet{arm1991} empirically proved that $W'$ is directly correlated
with [Fe/H].

Using 52 globular clusters \citet{rutledge1997b} found $\beta= 0.64\pm
0.02 {\rm \AA mag^{-1}}$ across the range $-2.1\la {\rm [Fe/H]}
\la-0.6$. \citet{cole2004} re-determined the value of $\beta$ for
their sample, which included globular and open clusters. They found
$\beta= 0.66\pm 0.03 {\rm \AA mag^{-1}}$ when using only globular
clusters, similar to the value found by \citet{rutledge1997b}, and
$\beta= 0.73\pm 0.04 {\rm \AA mag^{-1}}$ when including open clusters,
which covered an higher metallicity range ($-0.6\la {\rm [Fe/H]}
\la-0.15$).

The first uncertainty in using the CaT calibration is of course which
combination of the 3 CaT lines should be used. In the literature there
are several examples: \citet{rutledge1997a} used a weighted sum of the
3 lines; \citet[][hereafter T01]{tolstoy2001} excluded the first CaT
line from the sum; \citet[][hereafter C04]{cole2004} used an
unweighted sum of the 3 lines. Such a choice usually depends on
 the quality of the data-set: in the case of limited
signal-to-noise and with possible sky-line residual contamination, the
weakest line of the CaT is usually the least reliable and so the
determination of its EW is often doubtful.

These previous studies all calibrated the CaT $W'$ on the
\citet{carretta1997} scale (CG97), and \citet{friel2002} for the open
clusters, meaning that the relation between [Fe/H] and $W'$ has been
derived assuming that the calibration clusters have the metallicities
derived by CG97.  To define their [Fe/H] scale, CG97 re-analysed
high-quality EWs from different sources, using a homogeneous
compilation of stellar atmosphere parameters, $gf$ values and so on,
making the largest and most self-consistent analysis of this kind.  It
is important to note that even the HR values of the [Fe/H] for
globular clusters can differ significantly from each other
\citep[see][]{pritzl2005}. Obviously, choosing a different [Fe/H]
scale implies that the derived relations will be different.

This variety of approaches and calibrations can lead to a degree of
confusion when viewing the literature.  It is not possible to find out
a priori which way of summing CaT EWs and which calibration must be
used; we thus test each of the mentioned approaches from the
literature and see which one performs better for our globular clusters data-set.  In
addition, we derive our own calibration.

Each of the three calibrations that we examined from the literature
consists of three relations:

1) a linear combination of the CaT lines EW:
\begin{equation}
\label{eq:sumw_r97}
\Sigma W_{\rm R97}= 0.5 \, EW_1 + EW_2 + 0.6 \, EW_3
\end{equation}
\begin{equation}
\label{eq:sumw_t01}
\Sigma W_{\rm T01}= EW_2 + EW_3
\end{equation}
\begin{equation}
\label{eq:sumw_c04}
\Sigma W_{\rm C04}= EW_1 + EW_2 + EW_3
\end{equation}

2) a relation for the  ``reduced equivalent width'':
\begin{equation}
\label{eq:wp_r97}
W'_{\rm R97} = \Sigma W_{\rm R97} + 0.64 (\pm 0.02) \times (V - V_{\rm HB})
\end{equation}
\begin{equation}
\label{eq:wp_t01}
W'_{\rm T01} = \Sigma W_{\rm T01} + 0.64 (\pm 0.02) \times (V - V_{\rm HB})
\end{equation}
\begin{equation}
\label{eq:wp_c04}
W'_{\rm C04} = \Sigma W_{\rm C04} + 0.73 (\pm 0.04) \times (V - V_{\rm HB})
\end{equation}

3) the calibration of the ``reduced equivalent width'' with [Fe/H]:
\begin{equation}
\label{eq:feh_r97}
{\rm [Fe/H]}_{\rm CG97}^{\rm R97} = -2.66 (\pm 0.08) + 0.42 (\pm 0.02) \times W'_{\rm R97} 
\end{equation}
\begin{equation}
\label{eq:feh_t01}
{\rm [Fe/H]}_{\rm CG97}^{\rm T01} = -2.66 (\pm 0.08) + 0.42 (\pm 0.02) \times W'_{\rm T01} 
\end{equation}
\begin{equation}
\label{eq:feh_c04}
{\rm [Fe/H]}_{\rm CG97}^{\rm C04} = -2.966 (\pm 0.032) + 0.362 (\pm 0.014) \times W'_{\rm C04}
\end{equation}
where R97 stands for \citet{rutledge1997a, rutledge1997b}, T01 for \citet{tolstoy2001} and 
C04 for \citet{cole2004}. 

\begin{figure}
\begin{center}
\includegraphics[width=80mm]{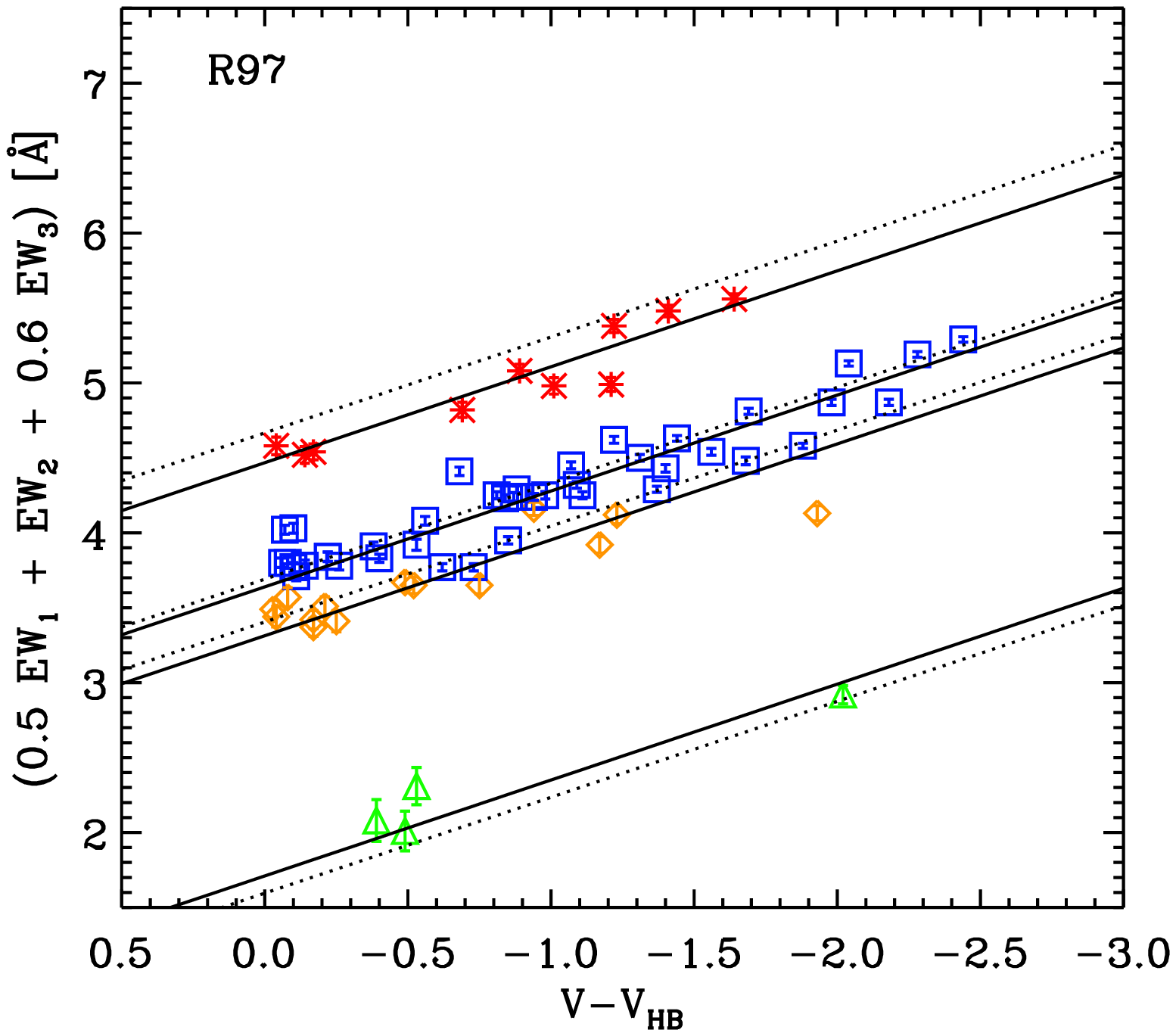}
\includegraphics[width=80mm]{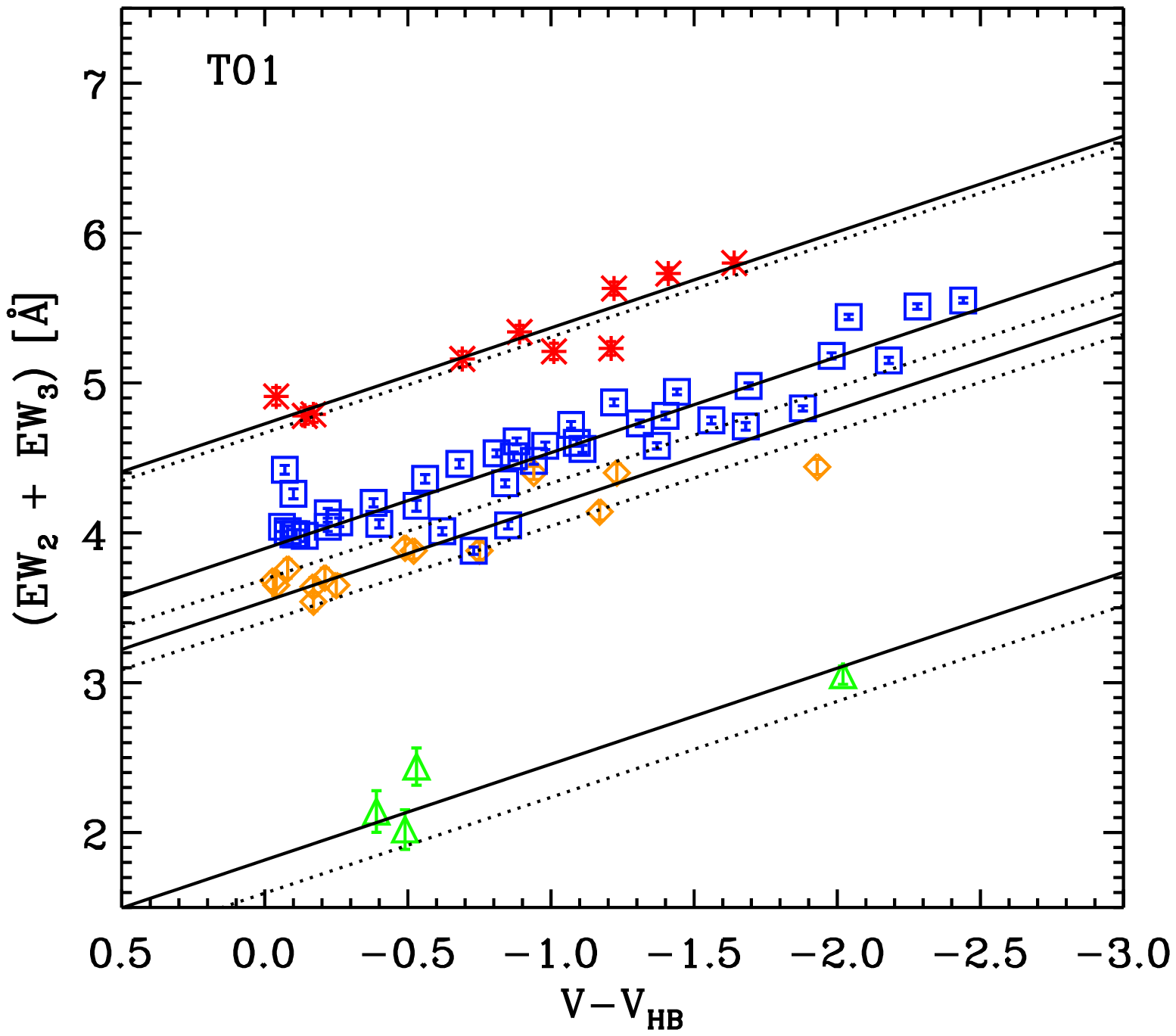}
\includegraphics[width=80mm]{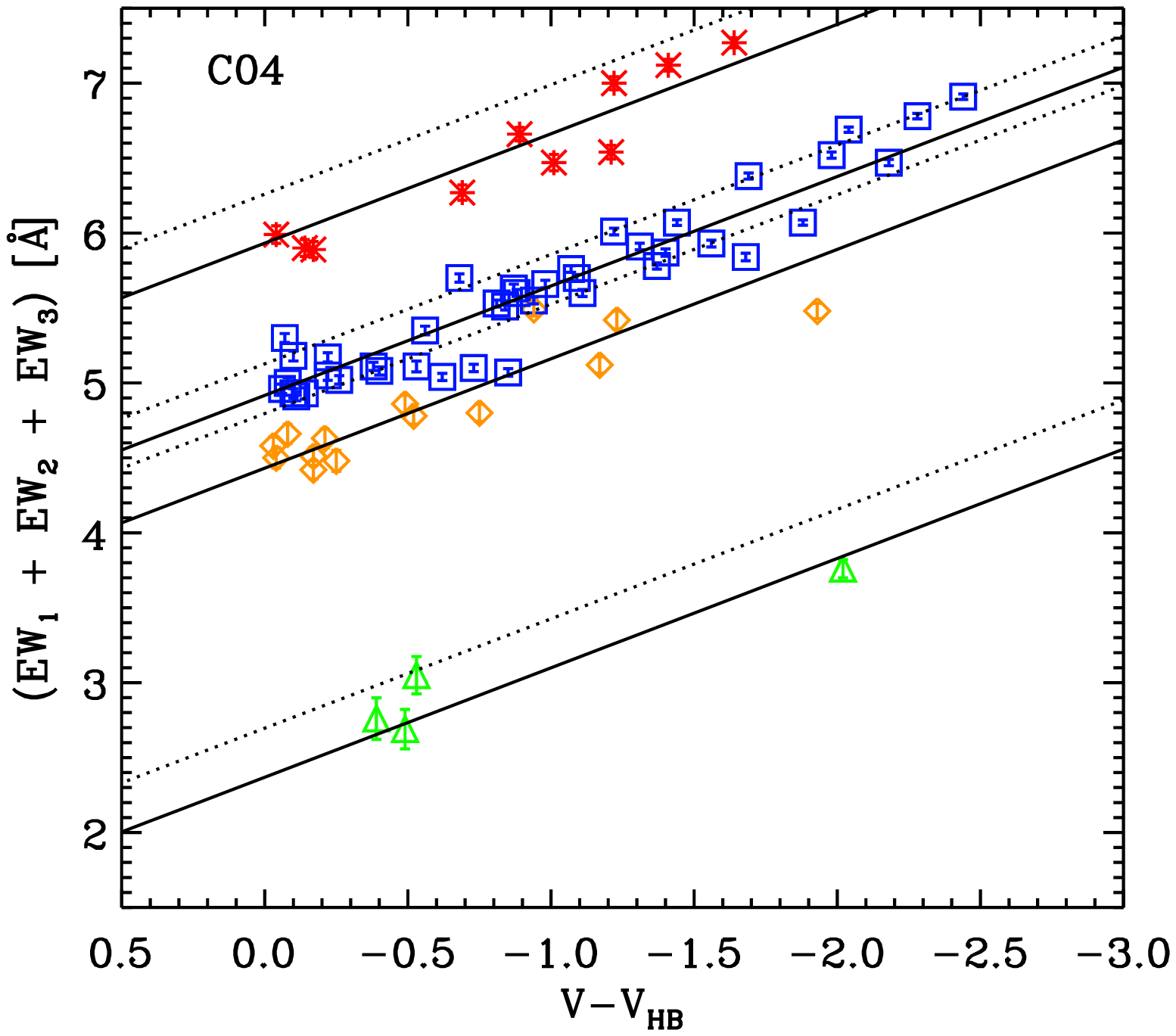}
\caption{The CaT calibrations using globular clusters: EWs vs. ($V-V_{\rm HB}$) for RGB stars  
in 4 globular clusters of different metallicities (asterisks: NGC104; squares: NGC5904; 
diamonds: NGC3201; triangles: NGC4590),  
using 3 different linear combinations of CaT lines (from top to bottom: R97, T01, C04). 
The dotted lines show the relations in R97, T01, C04, using the [Fe/H] published in CG97; the solid 
lines show our best-fitting relations, keeping the slope fixed within each calibration. 
The best-fitting metallicities are summarised in Table~\ref{tab:gc}. }
\label{fig:sumw_vvhb}
\end{center}
\end{figure}

Figure~\ref{fig:sumw_vvhb} shows $\Sigma W$ versus ($V-V_{\rm HB}$)
for the 4 globular clusters, summing the CaT lines as in Eqs.~(\ref{eq:sumw_r97}), (\ref{eq:sumw_t01}), (\ref{eq:sumw_c04}). As the
minimum S/N per \AA\, of these data is $\sim$ 40 and the median is
$\sim$ 100 per \AA\, the errors in the summed EW are very small, $\la
0.1$ \AA\,.  As consistency check we calculate the weighted average
slope for each of the calibrations and find that they are consistent
at the 1-$\sigma$ level with the previous works ($\beta_{\rm R97, this
work} = 0.59 \pm 0.04 \, {\rm \AA mag^{-1}} $, $\beta_{\rm T01, this
work} = 0.62 \pm 0.03 \, {\rm \AA mag^{-1}}$, $\beta_{\rm C04, this
work} = 0.79 \pm 0.04 \, {\rm \AA mag^{-1}}$).

\begin{figure}
\begin{center}
\includegraphics[width=60mm]{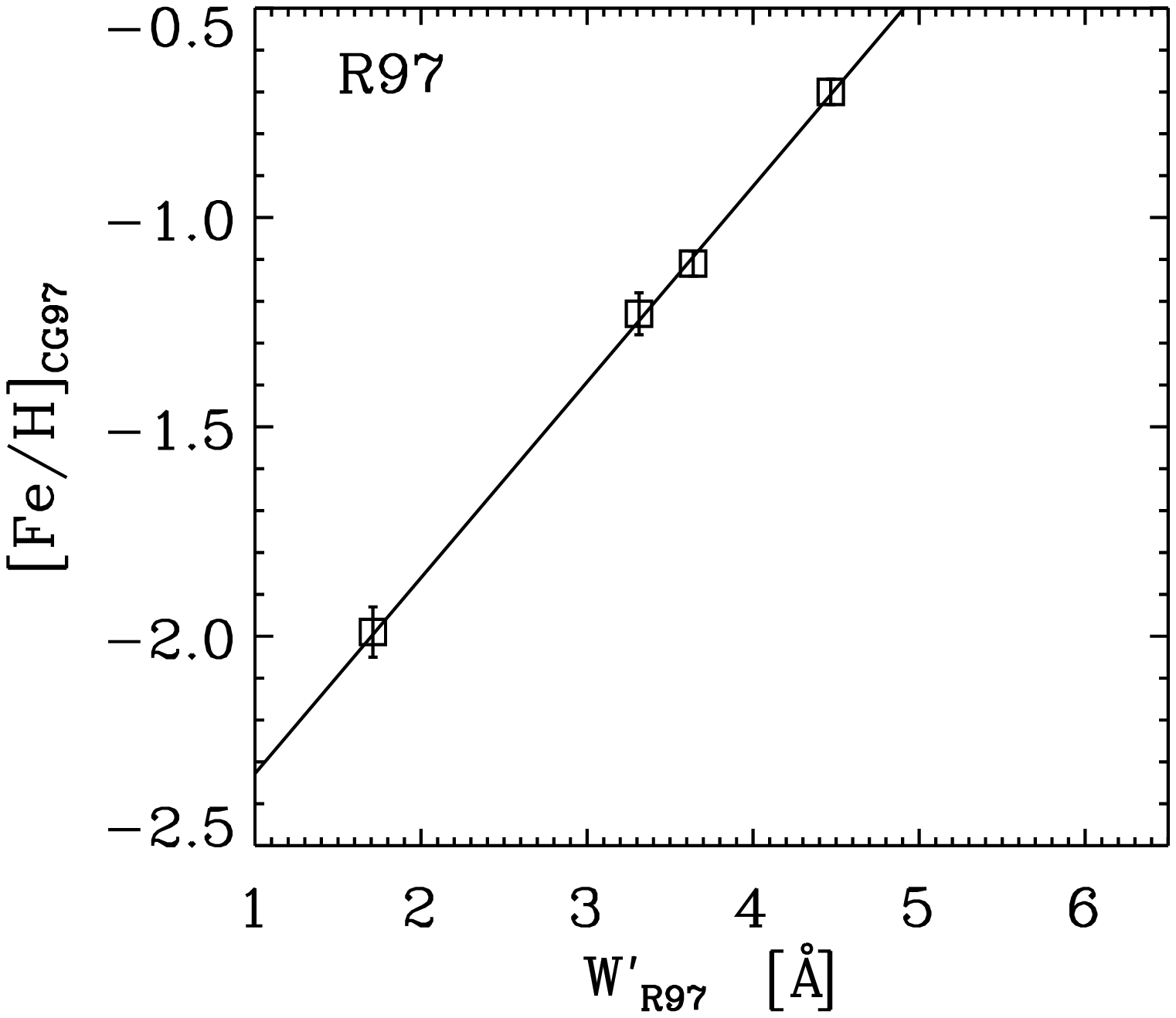}
\includegraphics[width=60mm]{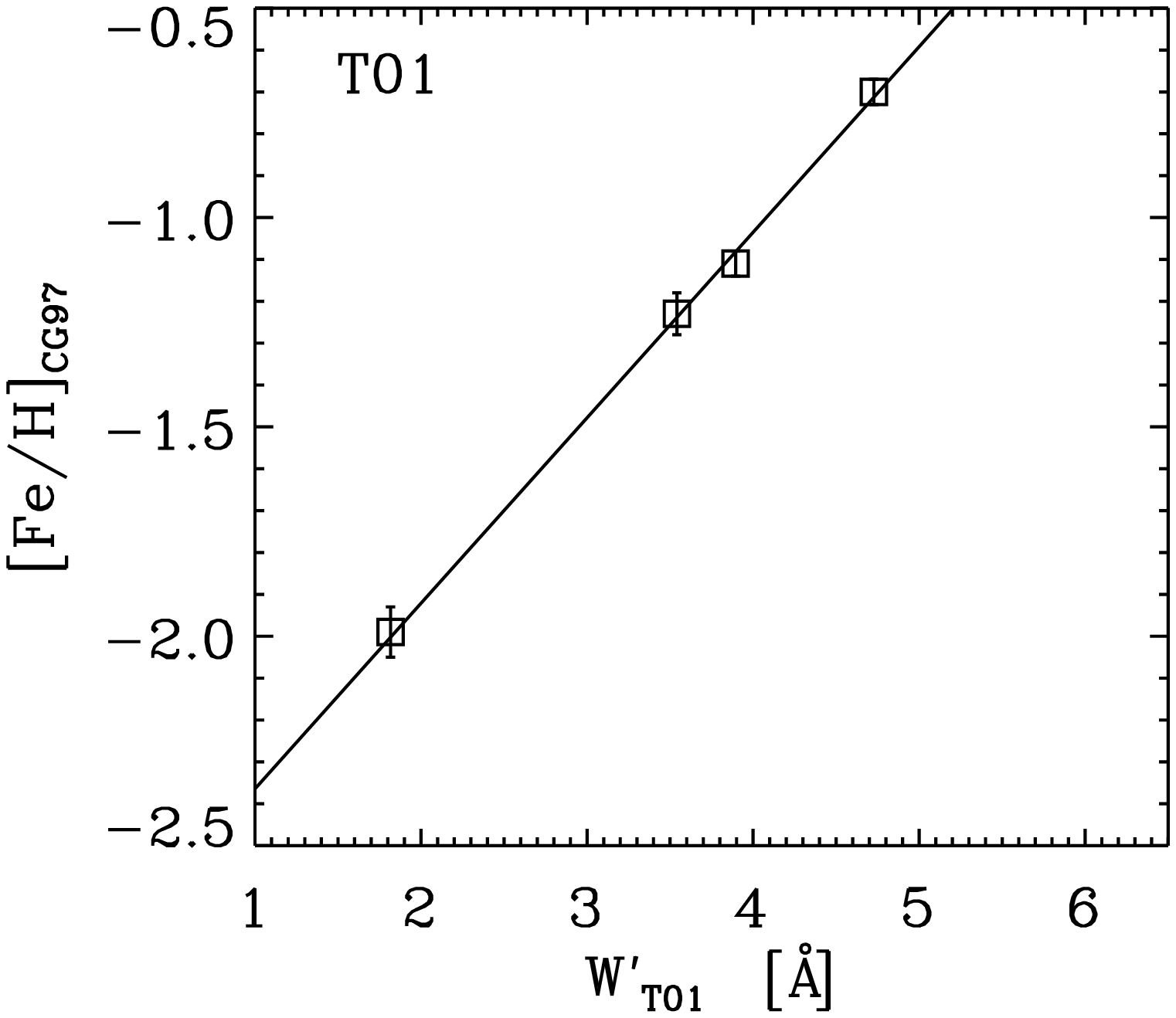}
\includegraphics[width=60mm]{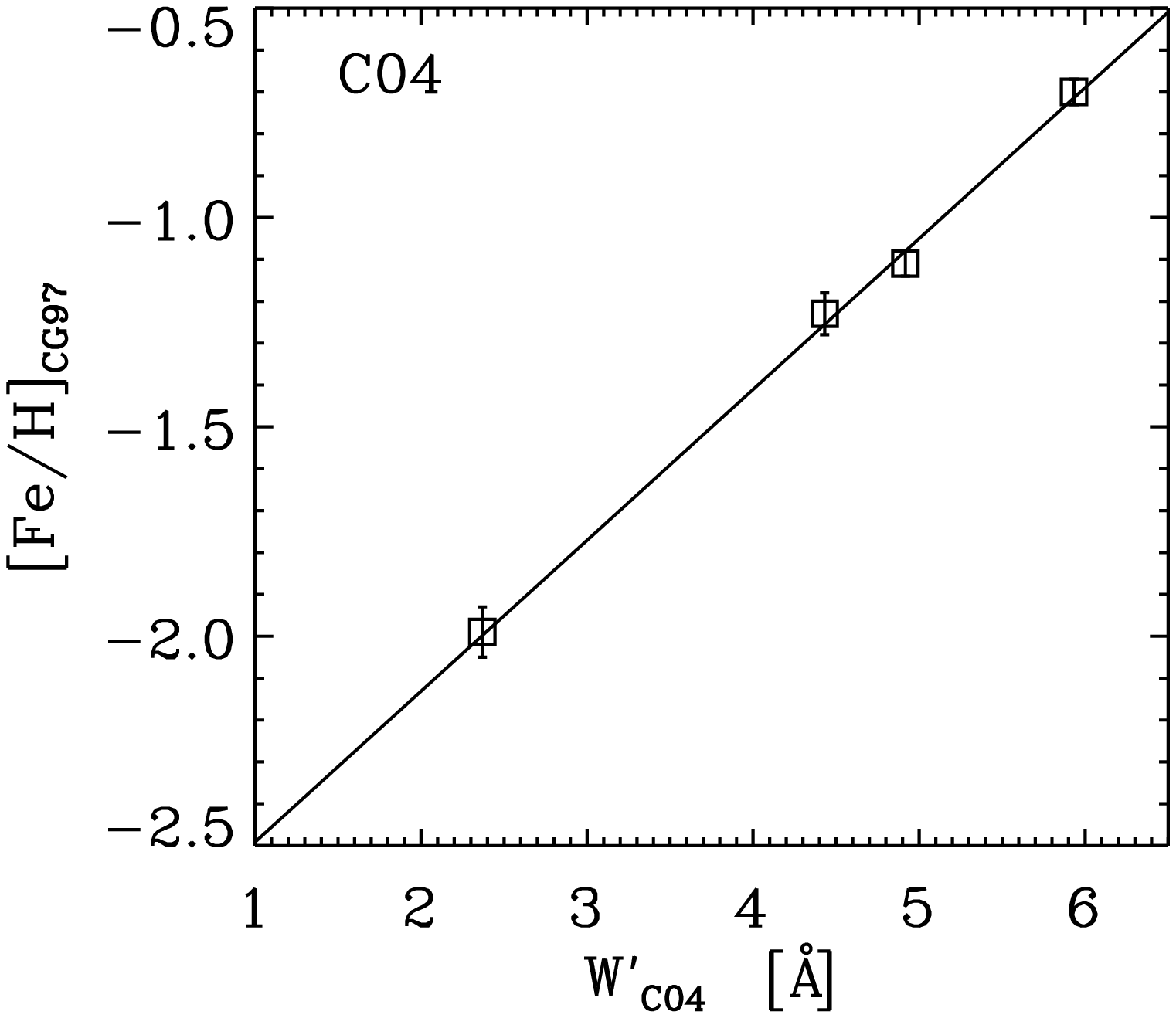}
\caption{Relation between CG97 [Fe/H] and $W'$ for the 4 calibrating 
clusters using R97, T01 and C04 calibrations (from top to bottom). The solid lines show 
the best-fitting CG97 [Fe/H] and $W'$ relation for each calibration.}
\label{fig:fe_wp}
\end{center}
\end{figure}

The metallicities derived from the best-fitting $W'$
 from an error-weighted linear fit of $\Sigma W$ versus ($V-V_{\rm
 HB}$) are summarised in
Table~\ref{tab:gc}. In general the metallicities derived from these
observations agree within 1$\sigma$ with the metallicities on the CG97
scale.  The metallicities from the C04 calibration seem to be
systematically lower by $\sim$ 0.1 dex.  The T01 calibration appears
to give the best performance. Thus, amongst the three relations in the
literature, we will apply the T01 to our dSphs dataset.
Figure~\ref{fig:fe_wp} shows that the relation between $W'$ and
[Fe/H]$_{\rm CG97}$ derived from the 4 globular clusters dataset is
linear in each case.

In order to derive our own calibration, we should repeat all the
steps, i.e. derive a $W'$ using the slope we find by fitting the
summed EW versus ($V-V_{\rm HB}$), and finding the best $W'$-[Fe/H]
relation. However, as mentioned before, the slopes we find are
consistent with those in the literature, and since the number of stars
in most of our calibration globular clusters is not large, we prefer
to use as relations those in the literature (Eqs.~\ref{eq:wp_r97}, \ref{eq:wp_t01}, \ref{eq:wp_c04}).  We
just repeat the last step and find the best-fitting $W'$-[Fe/H]
relations by performing an error-weighted linear-fit. The relations we
find are all consistent within 1-$\sigma$ with the relations in the
literature.  The calibration obtained using Eq.~(\ref{eq:wp_t01}) appears to give
the best results and is the following:
\begin{equation}
\label{eq:feh_b07}
{\rm [Fe/H]}_{\rm CG97}^{\rm this work}= (-2.81 \pm 0.16) + (0.44 \pm 0.04) \times W'_{\rm T01}  
\end{equation} 

In the following section we test to see if the [Fe/H]-$W'$ relation
derived for the globular cluster sample is reliable when applied to
RGB field stars in galaxies. To do so, we apply Eqs.~(\ref{eq:feh_t01}) and (\ref{eq:feh_b07}) to
the Sculptor and Fornax sample and we compare the metallicities so
derived to the HR metallicities.

\section{Comparison to High Resolution metallicity measurements} \label{sec:calibration_hr}
One major uncertainty in applying the CaT method to field stars in galaxies, 
for example in dSphs, is that the [Fe/H]-CaT $W'$ relations have so 
far been calibrated exclusively on globular clusters. Some dSphs contain 
intermediate age and even young stellar populations and have a large spread in 
metallicity, whilst the above relations have been derived for single age 
stellar population older than 10 Gyr, over a relatively narrow metallicity 
range and also a very narrow range in [$\alpha$/Fe], which is very different from 
composite stellar populations. Furthermore, in composite populations it is 
more difficult to assign a unique magnitude for the HB, although
\citet{cole2000, cole2004} showed that the uncertainty due to this effect is 
$\sim$ 0.05 dex, which is not significant compared to the intrinsic precision of the 
method ($\sim$ 0.1 dex).

The only way to reliably test the standard globular cluster calibration for 
dSph field stars is to compare the [Fe/H] derived from CaT EWs to that 
obtained from direct measurements in HR 
observations of the same field stars. HR metallicities should be more accurate than 
CaT measurements because the iron abundance is not inferred from other elements
but obtained by direct measurement of typically more than 60 separate Fe lines 
with two different ionisation states (i.e. Fe\,{\small I} and Fe\,{\small II}). 
  
Until now a comparison between HR and LR [Fe/H] has not been thoroughly made 
for dSph field stars, partly due to the lack of a large sample of overlapping measurements.  
Thanks to 
instruments like FLAMES, it is now possible to get suitable comparison 
spectra for many objects at the same time. 

As part of the DART Large Program at ESO, HR FLAMES spectra have been taken 
for 93 probable Sculptor velocity member stars in the central regions of the 
Sculptor dSph (Hill et al. 2008, in preparation) for which there is also LR CaT data 
\citep{tolstoy2004}. A similar study was made of a central field 
in Fornax (Letarte et al. 2008, in preparation; Letarte 2007), for which 36 stars overlap our 
LR sample \citep{battaglia2006}.

These observations consist of R$\sim$20000 
resolution spectra of $\sim$80 stars in the centre of both Fornax and Sculptor,
covering $\sim$60nm in three different FLAMES set-ups (534-562nm;
612-641nm; 631-670nm), and reaching typical S/N of 30 per 0.05nm
pixel. The chemical analysis of the sample was performed using OSMARCS
one-dimensional stellar atmosphere models in LTE \citep[OSMARCS models,][]{gustafsson2003, gustafsson2007}, 
an extension of the OSMARCS models referenced above (Pletz, private communication), 
 and a standard EW
analysis. Stellar parameters were determined using a combination of
photometric indices (V, I, J, H, K) and spectroscopic indicators
(excitation and ionisation equilibrium). The results include the
abundances of $\sim$10 elements, including iron and calcium which are
reported here for comparison to the LR results (see Table~\ref{tab:results_dsphs}). Error bars on HR abundances 
indicated on the plots refer to the combined abundance
measurement errors and propagated stellar parameters uncertainties.

The detailed description of the data reduction and analysis of the
HR spectroscopic data can be found in a
series of papers (Hill et al. 2008; Letarte et al. 2008; Letarte 2007). 
Also note that the HR results have been put on to UVES system \citep[e.g.,][]{letarte2007}. 

\begin{itemize}
\item {\bf CaT calibration using HR data:} 
First we determine the relation between [Fe/H]$_{\rm HR}$ and CaT $W'$
directly for the dSph data. We assumed $V_{HB}=20.13$ for Sculptor and $V_{HB}=21.29$ for Fornax,
taken from \citet{IH1995}. Figure~\ref{fig:wp_hr} shows the HR [Fe/H]
of the overlapping stars between the HR and LR samples plotted against
their reduced CaT equivalent width (Eq.~\ref{eq:wp_t01}). The best linear fit we
obtain allowing for errors in both coordinates is
\begin{equation}
\label{eq:wpfeh_fnxscl}
{\rm [Fe/H]}_{\rm HR}= -2.94 (\pm 0.04) + 0.49 (\pm 0.01) W'
\end{equation}
This calibration is consistent at the 1-$\sigma$ level with the
calibration derived in the previous section for 4 globular clusters,
however there are some differences. For small $W'$ the above relation
predicts a [Fe/H] $\sim$ 0.1 dex lower than from the globular cluster
calibration given in Eq.~(\ref{eq:feh_b07}), whilst the opposite happens for large
$W'$.

\item {\bf Comparison between HR and LR results using FLAMES globular
      cluster calibration:}
The traditional globular cluster calibration of CaT is now applied to
our LR CaT $W'$ and the results compared to [Fe/H]$_{\rm HR}$. 
Figure~\ref{fig:met_newcal} shows the comparison between HR [Fe/H] and
CaT [Fe/H] from our globular cluster calibration (Eq.~\ref{eq:feh_b07}). As also
indicated in the previous figure, the two methods are generally in
good agreement.  The average difference is $\Delta \rm [Fe/H]=
\rm{[Fe/H]_{\rm LR} - [Fe/H]_{\rm HR}}= -0.04 \pm 0.02$ dex and the
spread is 0.17$\pm$0.02 dex, which is comparable with the measurement
errors.  We can thus apply the globular cluster calibration to dSph
field stars with some confidence, between $-2.5 \la$[Fe/H]$\la - 0.8$
dex, where the relation between HR and LR data is linear. It is
unclear if the comparison at [Fe/H]$> - 0.8$ dex indicates a
non-linearity in the relationship, as the appearance of
``non-linearity'' is given by $\sim$ 4 stars. To make a concrete
statement requires more data at high [Fe/H].

The average difference calculated for the entire sample would suggest
the absence of systematics in our evaluation of [Fe/H]$_{\rm
LR}$. However when plotting $\Delta \rm [Fe/H]$ versus [Fe/H]$_{\rm
HR}$ (bottom panel Fig.~\ref{fig:met_newcal}) a trend is visible, such
that the LR [Fe/H] is overestimated of $\sim$ 0.1 dex at [Fe/H]$_{\rm
HR} \la -2.2$, and underestimated of $\sim$0.1-0.2 dex at the high
[Fe/H] end, at [Fe/H]$_{\rm HR} \ga -1.2$, which instead suggests the
presence of systematics in the [Fe/H]$_{\rm LR}$ derivation.

\begin{figure}
\begin{center}
\includegraphics[width=80mm]{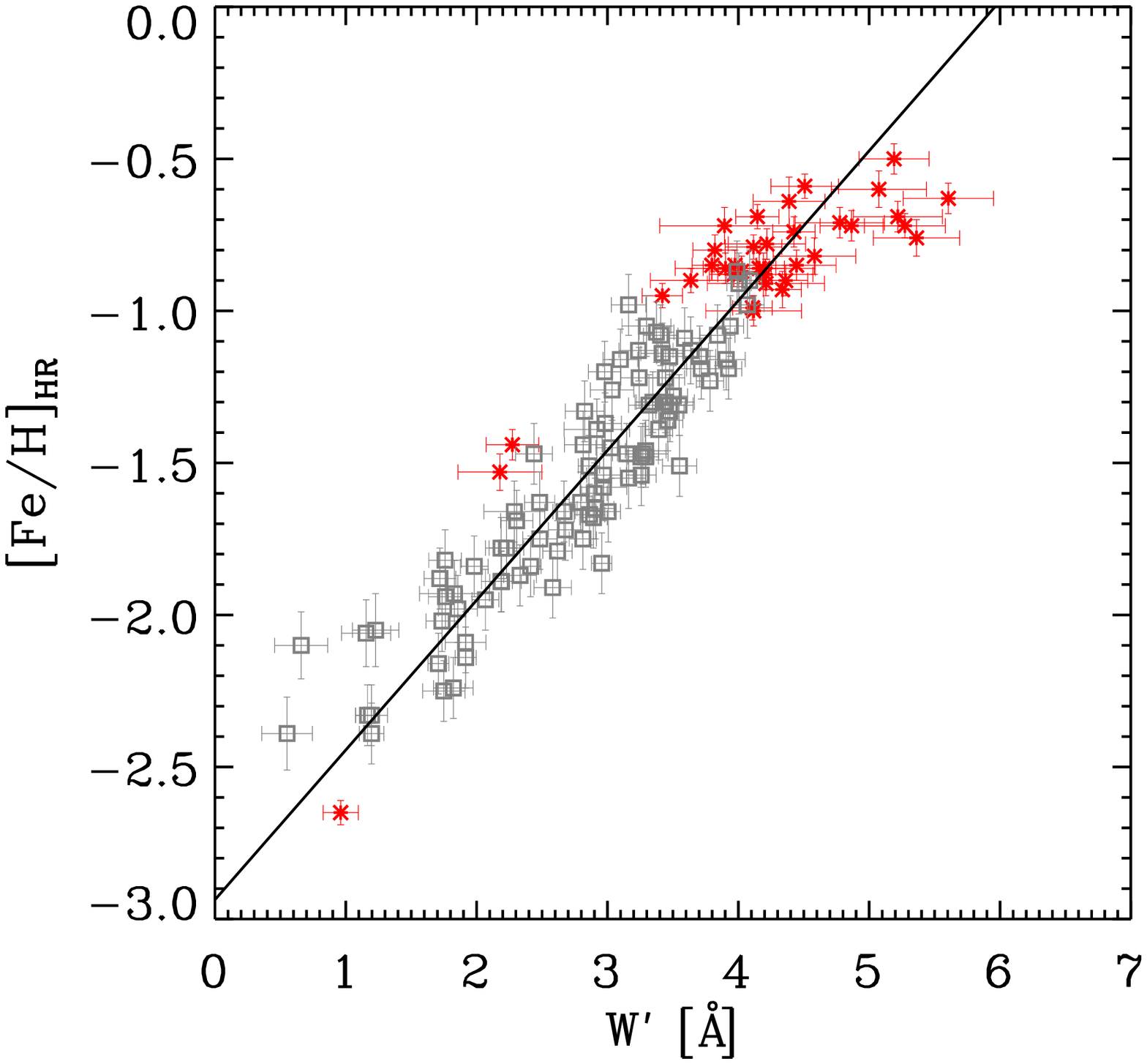}
\includegraphics[width=80mm]{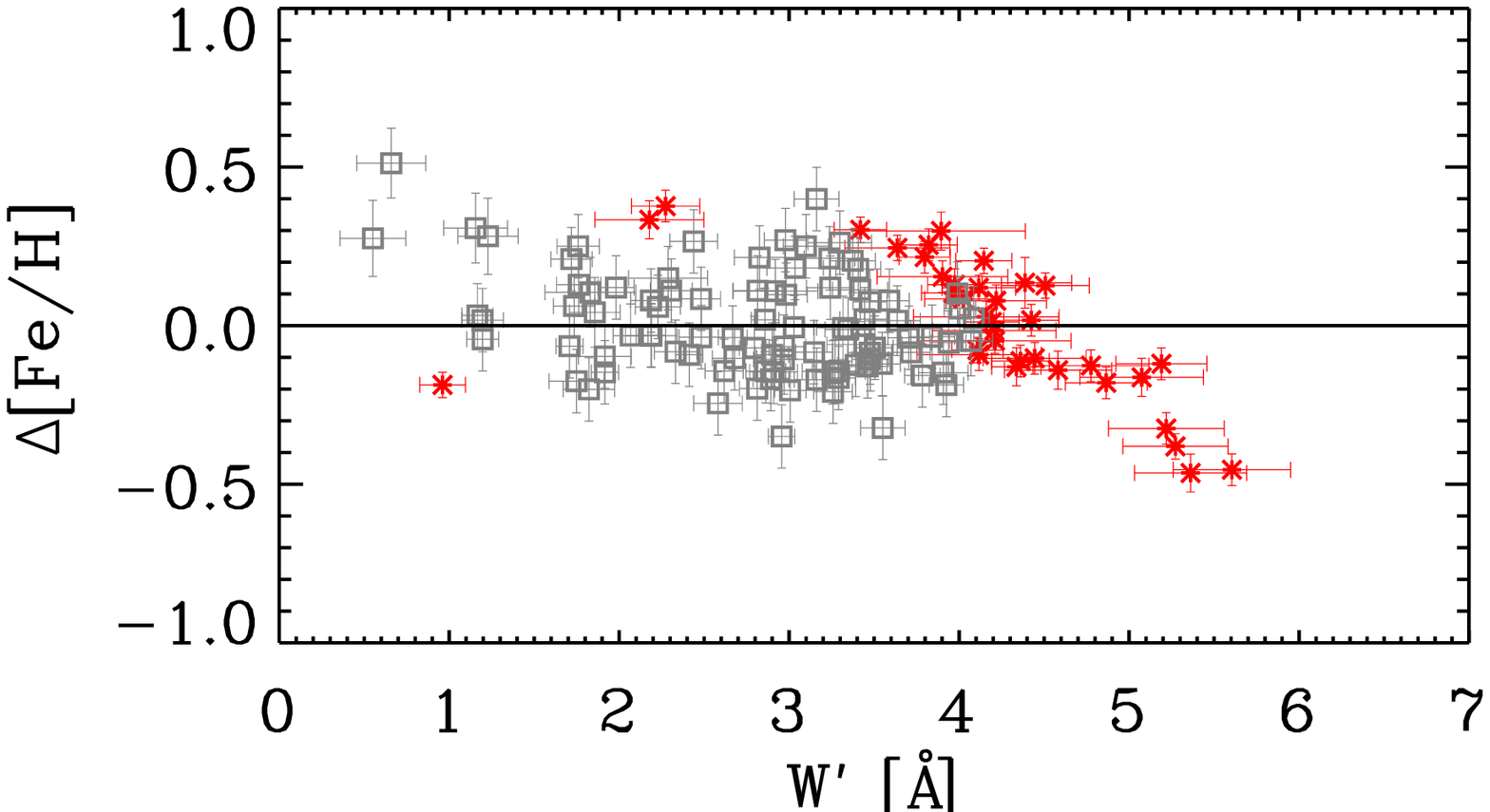}
\caption{Top: HR [Fe/H] versus CaT $W'$ for the RGB stars 
overlapping between the HR and LR sample for the Sculptor and Fornax dSphs (93 in Sculptor, squares, and 36 stars in Fornax, asterisks). 
The solid line represents the best error-weighted linear fit to the data 
($-2.94 (\pm 0.04) + 0.49 (\pm 0.01) W'$). 
Bottom: Residuals to the linear fit. The 1$\sigma$ $y-$errorbars are the errors in HR [Fe/H]. 
The r.m.s. around the relation is 0.19 dex (MAD = 0.12 dex), comparable to the measurement errors.}
\label{fig:wp_hr}
\end{center}
\end{figure} 

\begin{figure}
\begin{center}
\includegraphics[width=80mm]{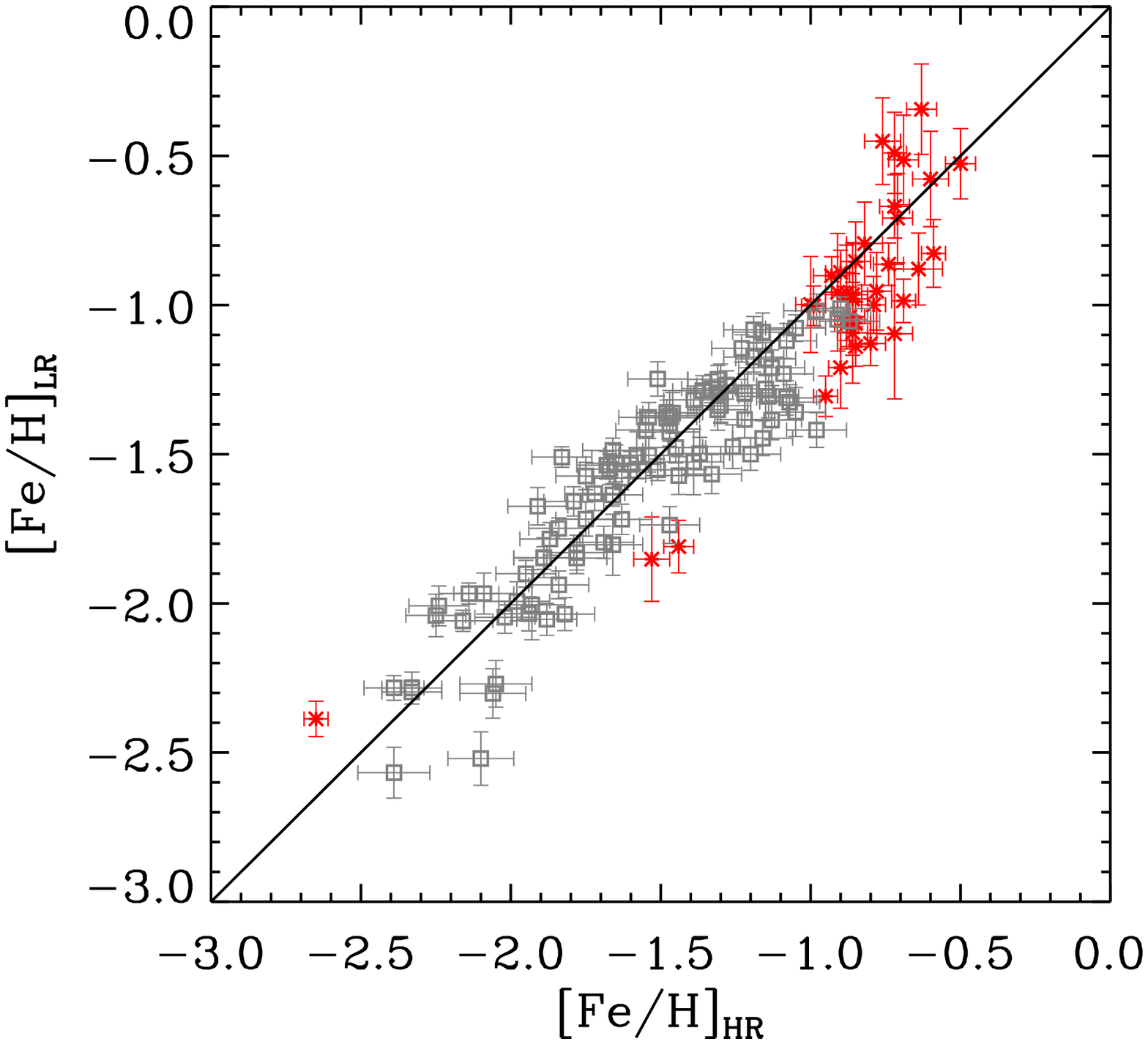}
\includegraphics[width=80mm]{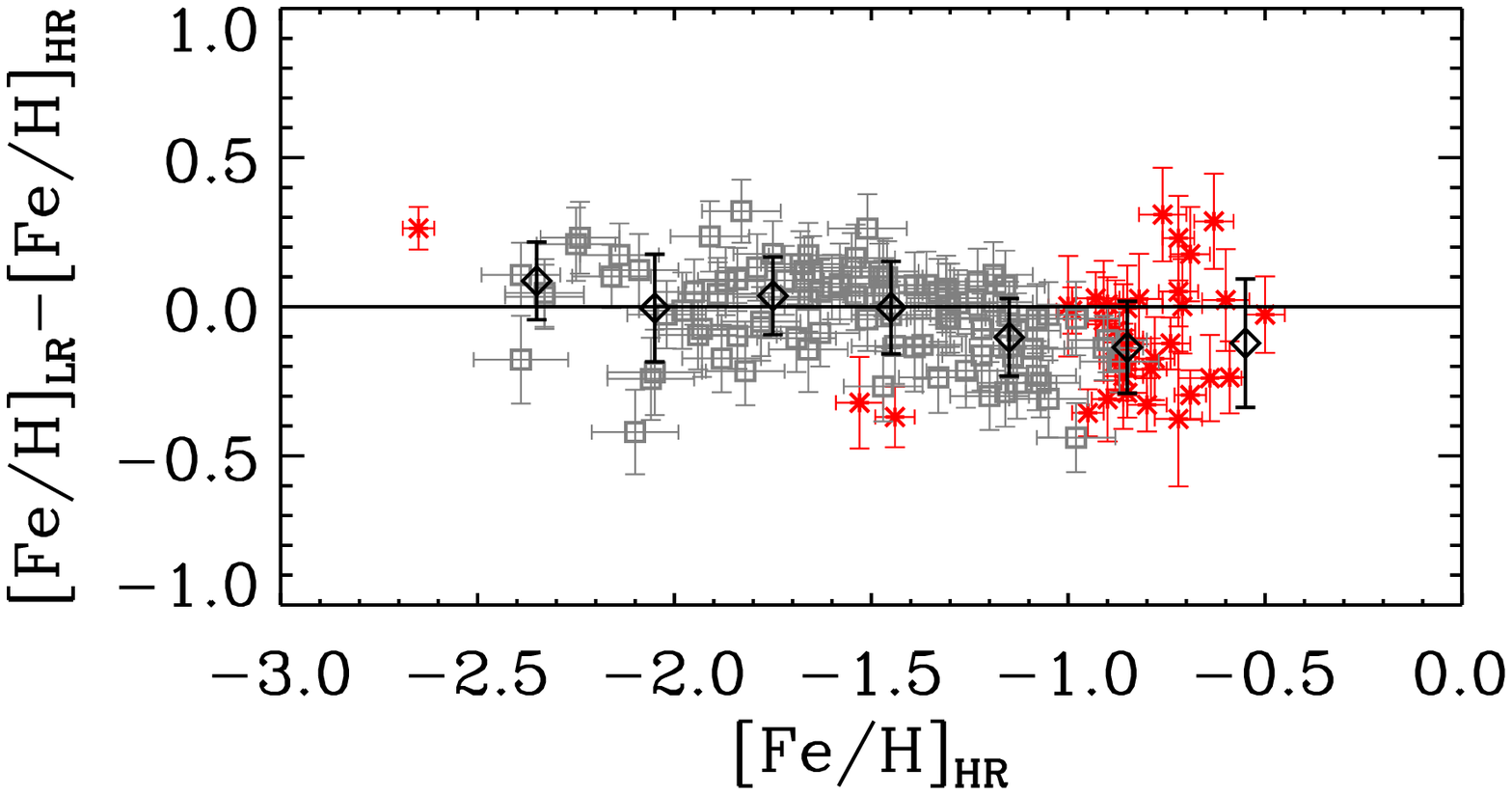}
\caption{Top: Comparison between HR and CaT [Fe/H] for the Sculptor (squares) and Fornax (asterisks) dSphs. 
The solid line indicates the one-to-one relation. The CaT metallicities are derived using Eq.~(\ref{eq:feh_b07}). 
Bottom: The above comparison plotted as a difference. The diamonds with errorbars show 
the weighted average (and dispersion) of [Fe/H]$_{\rm LR}-$[Fe/H]$_{\rm HR}$ in bins of 0.3 dex. 
The average difference is [Fe/H]$_{\rm LR}-$[Fe/H]$_{\rm HR}= -0.04 \pm 0.02$ 
dex. The r.m.s. scatter around the one-to-one relation 
is 0.17 dex (MAD = 0.12 dex). For the stars with [Fe/H]$_{\rm HR} > -0.8$ the scatter 
and MAD are 0.2 dex and 0.21 dex, respectively.}
\label{fig:met_newcal}
\end{center}
\end{figure}

\begin{figure}
\begin{center}
\includegraphics[width=80mm]{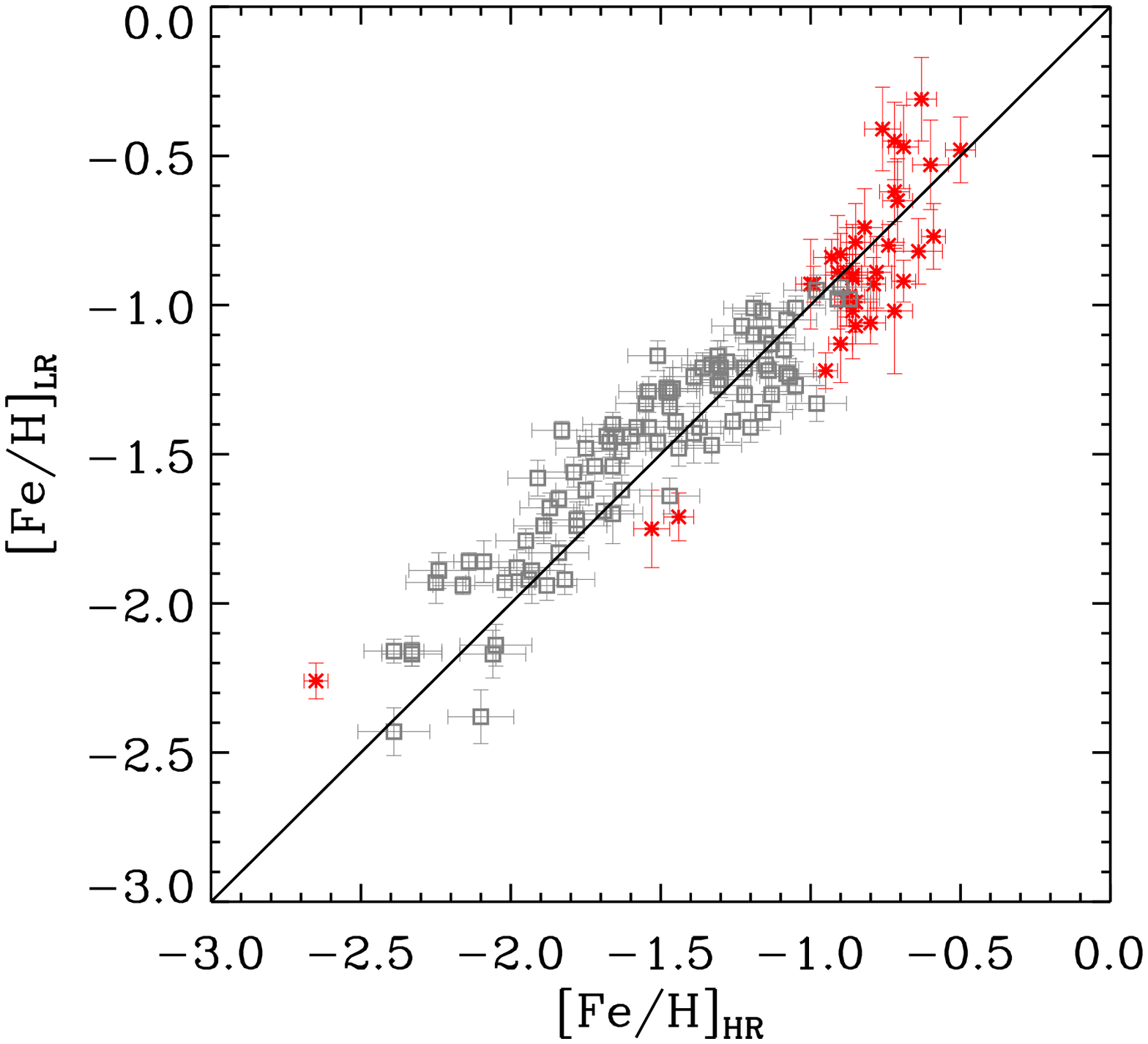}
\includegraphics[width=80mm]{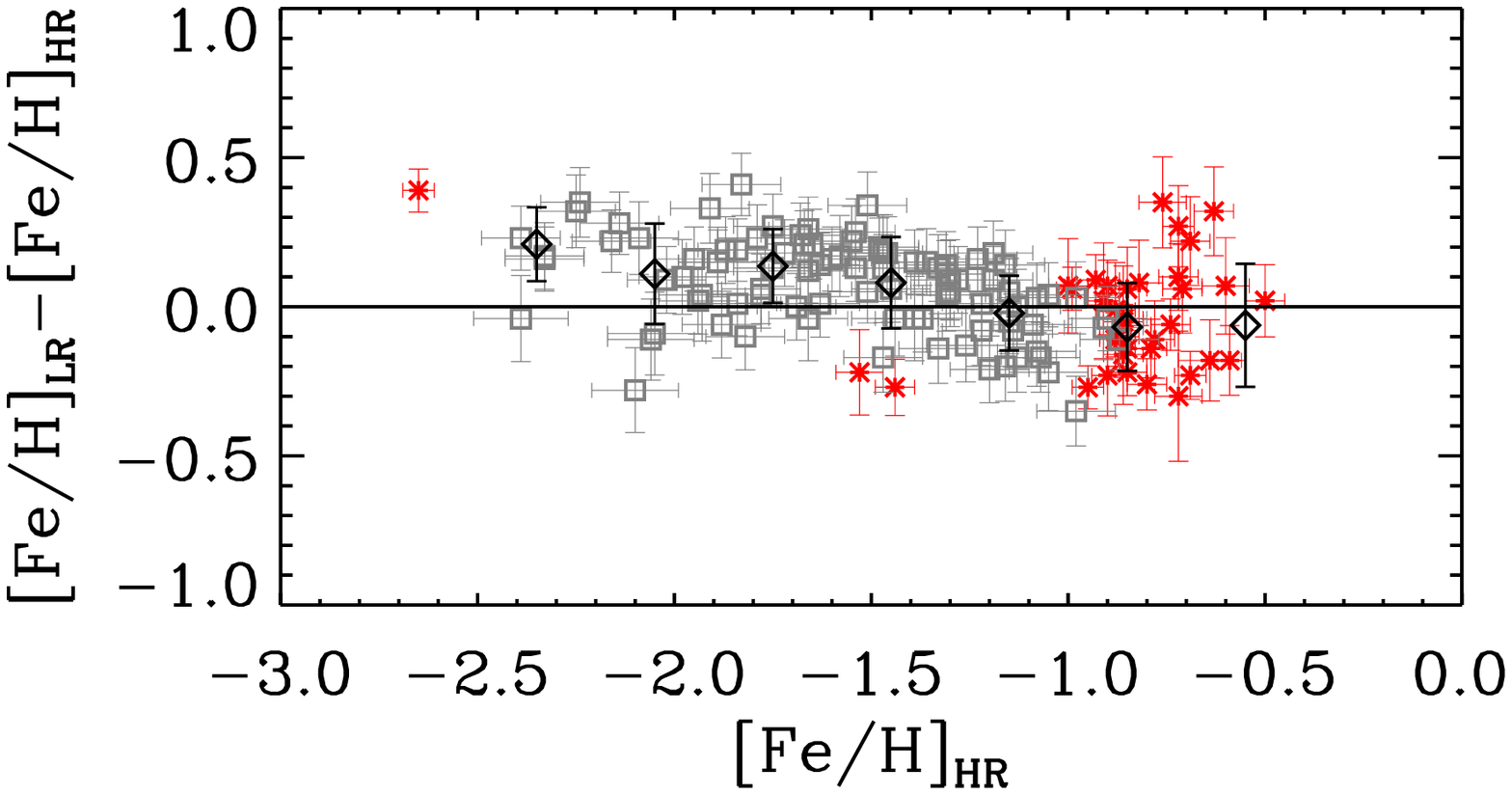}
\caption{Top: As previous figure, but the CaT metallicities are derived from Eq.~(\ref{eq:feh_t01}). 
Though the average 
difference is very small, [Fe/H]$_{\rm LR}-$[Fe/H]$_{\rm HR}=$0.04$\pm$0.02 dex, 
the bottom panel shows that there is a gradient in [Fe/H]$_{\rm LR}-$[Fe/H]$_{\rm HR}$ 
with [Fe/H], more enhanced than in the previous case. 
The r.m.s. scatter around the one-to-one relation 
is 0.17 dex (MAD = 0.12 dex).}
\label{fig:met_tolstoycal}
\end{center}
\end{figure} 

\item {\bf Comparison between HR and LR results using globular cluster calibrations from the literature:} 
As a comparison Fig.~\ref{fig:met_tolstoycal} shows the CaT [Fe/H] derived from the 
T01 calibration (Eq.~\ref{eq:feh_t01}), which we used in \citet{tolstoy2004}, 
\citet{battaglia2006} and \citet{helmi2006}. The trend with [Fe/H] is similar to the previous case, 
because of the similar slopes in the [Fe/H] versus $W'$ relations.  
The systematic shifts are more enhanced with respect to the globular cluster calibration 
at the low [Fe/H] end, whilst the performance at the high [Fe/H] end is 
better than for Eq.~(\ref{eq:feh_b07}). 
\end{itemize}
Both the relations in the literature and our FLAMES globular 
clusters calibration are in agreement with the HR data in the range $-2.5 \la$[Fe/H]$\la - 0.8$ dex. 
Given the slight differences between the relations, the choice of one over the other does not 
affect considerably the derived metallicities. 
In the remainder of the paper we follow
 the standard approach and use our own globular cluster calibration (Eq.~\ref{eq:feh_b07}). 
We also checked that [Fe/H] derived from our CaT calibration is in agreement with 
other derivations, such as in \citet{pont2004}.

\section{Possible sources of uncertainty in the CaT metallicity determination} \label{sec:uncert}
Although [Fe/H] derived from the CaT method is in good agreement 
with the derivation from HR measurements, it would be interesting to understand the causes 
of the apparent systematics.  

As mentioned in Sect.~\ref{sec:calibration_hr}, the traditional CaT calibration has a number of implicit 
assumptions in it: 1) [Ca/H] does not play a role in determining $W'$, and thus 
one can exclude it from the calibration; 
2) alternatively, if it plays a role, one assumes that globular clusters and dSphs have similar [Ca/Fe]; 
3) the CaT EW is a better estimator of [Fe/H] than [Ca/H]; 
4) the effect of age differences between dSphs and globular clusters stars can be neglected. 
In the following we explore the validity of these assumptions.

\subsection{CaT EW as [Ca/H] estimator}
\subsubsection{Calibration from globular clusters}
To test assumption 3) we derive a relation 
between $W'$ and [Ca/H] using the globular clusters as calibrators, and we apply it to our sample of dSphs. 
If $W'$ traces [Ca/H], then we should expect a good one-to-one relation between the Ca abundance derived 
from the globular clusters calibration and the Ca abundance derived from the direct HR measurements. 

In order to derive the [Ca/H] values needed for the globular cluster $W'$-[Ca/H] relation 
we assumed an average value of [Ca/Fe]=0.235 dex for the 4 globular clusters 
(see Table~\ref{tab:gc} for the individual values), and we used [Ca/H]$=$[Ca/Fe]+[Fe/H], where 
[Fe/H] are the individual values from CG97 (see Table~\ref{tab:gc}).

\begin{figure}
\begin{center}
\includegraphics[width=70mm]{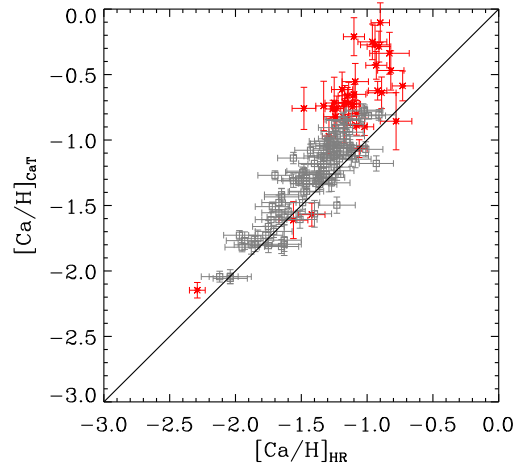}
\caption{[Ca/H]$_{\rm CaT}$ from Eq.~(\ref{eq:cah_gcs}) versus [Ca/H]$_{\rm HR}$ 
for the Sculptor (squares) and Fornax (asterisks) dSphs. The solid line shows the 
one-to-one relation.
}
\label{fig:cah}
\end{center}
\end{figure}

The best-fitting relation we obtain from the globular clusters is
\begin{equation}
\label{eq:cah_gcs}
\rm [Ca/H]_{CaT} =  -2.57 (\pm 0.18) + 0.44 (\pm 0.05) dex/\AA \ {\it W'}.
\end{equation}
Using the individual [Ca/Fe] values (instead of the average [Ca/Fe]=0.235 dex) does not change the best-fitting 
relation, it just increases the resulting minimum $\chi^2$ value of the fit. 
Figure~\ref{fig:cah} shows the comparison between [Ca/H]$_{\rm CaT}$ and the corresponding [Ca/H] derived 
from HR measurements for the Sculptor and Fornax dSphs\footnote{Out of the 93 stars in the overlapping sample in Sculptor, 
4 had too low signal-to-noise to allow a determination of the HR [Ca/H]}. The agreement between the two 
is not as good as for [Fe/H]: the scatter is much larger and [Ca/H]$_{\rm CaT}$ is in general 
overestimated, especially at [Ca/H]$\ga -1.3$. The comparison gets worse if we approximate the trend of [Ca/Fe] with [Fe/H]
with a function, to circumvent the effect of the large errors in [Ca/Fe]. 

Thus the Ca 
abundance cannot be the only factor that drives the $W'$. 

\subsubsection{Calibration from the literature}
We compared our results to those of \citet{bosler2006}. 
They used 15 Galactic globular 
clusters and 6 open clusters, 
covering the range 2 Gyr $\la$ age $\la$ 14 Gyr and $-2.2 <$ [Ca/H] $< +0.2$, 
to derive a calibration between CaT $W'$ and Ca abundance, the idea 
being that using $W'$ as tracer of [Ca/H] instead of [Fe/H] one can apply the calibration to 
systems such as globular cluster or dSphs, without worrying about the different [Ca/Fe] trends. 
The relation they derive is:
\begin{equation}
\rm [Ca/H]_{Bosler} = -2.778(\pm 0.061) + 0.470 (\pm 0.016) dex/\AA \ {\it W'}
\end{equation}
and they apply it to the Leo~I and Leo~II dSphs. This relation is consistent with our Eq.~(\ref{eq:cah_gcs}), 
and this results in a poor 
comparison between [Ca/H]$_{\rm CaT}$ and [Ca/H]$_{\rm HR}$.

\subsection{Exploring the effect of [Ca/Fe]}
\begin{figure}
\begin{center}
\includegraphics[width=80mm]{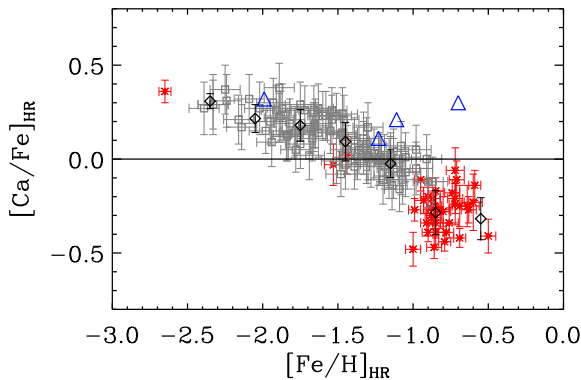}
\caption{HR [Ca/Fe] versus HR [Fe/H] for the Sculptor (squares) and 
Fornax (asterisks) dSphs. The triangles show the values for the calibration 
globular clusters (Table~\ref{tab:gc}). The diamonds with errorbars show a weighted 
average and dispersion of [Ca/Fe] in [Fe/H] bins.}
\label{fig:cafe}
\end{center}
\end{figure}

Another of the uncertainties in applying the CaT method to composite
stellar populations (i.e. galaxies) is the varied and extended star
formation histories of these systems, which results in a range of
[Ca/Fe] values and a trend with [Fe/H].  Figure~\ref{fig:cafe} shows
that our calibrating globular clusters have an almost constant [Ca/Fe]
trend with [Fe/H] (except for NGC3201), whilst both for the Sculptor
and Fornax dSph [Ca/Fe] decreases with [Fe/H].  At [Fe/H]$\la -1.1$
there is overlap between the [Ca/Fe] values for the globular clusters
and the dSphs, although [Ca/Fe] in the globular clusters is larger
than the average for the dSphs at the same [Fe/H]; at [Fe/H]$\sim
-1.1$ the [Ca/Fe] value for the globular clusters is $\sim$ 0.2 dex
larger than the average value for Sculptor and at [Fe/H]$=-0.7$ is
$\sim$ 0.5 larger than the average value for the Fornax dSph.  This
results in a smaller [Ca/H] abundance for globular clusters with
respect to our dSphs sample at the low [Fe/H] end, and we have the
opposite effect at the high [Fe/H] end.  If the $W'$ is an increasing
function of both [Fe/H] and [Ca/H] abundances, to neglect the effect
of [Ca/H] in our globular cluster calibration would result in an
overestimated [Fe/H]$_{\rm LR}$ with respect to [Fe/H]$_{\rm HR}$ in
the region where [Ca/H]$_{\rm dSph} > $ [Ca/H]$_{\rm GC}$; instead we
would underestimate [Fe/H]$_{\rm LR}$ with respect to [Fe/H]$_{\rm HR}$ 
in the region where [Ca/H]$_{\rm dSph} < $ [Ca/H]$_{\rm GC}$.
This goes in the same direction to what we see in
Fig.~\ref{fig:met_newcal}.  This suggests that [Ca/H] might also play
a role in determining the CaT $W'$.

\begin{figure}
\begin{center}
\includegraphics[width=60mm]{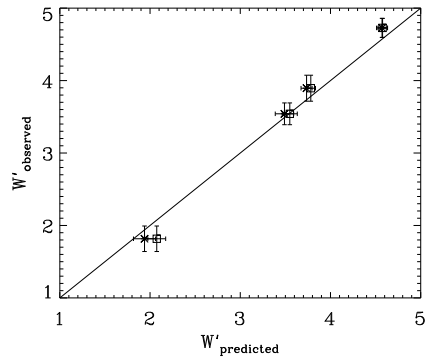}
\caption{Observed CaT $W'$ for our globular cluster sample versus CaT $W'$ 
derived from a linear combination of Fe and Ca abundances (Eq.~\ref{eq:wp_comb}), 
calibrated on the dSphs sample (squares) and from the linear dependence 
on Fe abundance given by Eq.~(\ref{eq:wpfeh_fnxscl}) (asterisks). 
The solid line indicates the one-to-one relation.}
\label{fig:wp_cahfeh}
\end{center}
\end{figure}

It is not obvious how to quantify this effect as it is not known how
the CaT $W'$ depends on [Ca/H] (linearly, quadratically, etc). We
tested for a simple dependence of this kind:
\begin{equation}
\label{eq:wp_comb}
W' = a {\rm [Fe/H]} + b {\rm [Ca/H]} + c
\end{equation}
where $a, b$ and $c$ are constants. We derive $a= 1.58 \pm 0.09$, $b=0.36 \pm 0.13$,
$c=5.85 \pm 0.07$ by fitting the observed values of $W'$ from the dSphs dataset
as a function of the corresponding [Fe/H] and [Ca/H] from HR
measurements. This relation, which suggests that [Ca/H] does have an impact on $W'$, but it is
less significant than [Fe/H], agrees well with the observed values of
$W'$ when applied to the globular cluster data
(Fig.~\ref{fig:wp_cahfeh}), although it does not remove the trend altogether. 
The importance of [Ca/H] in driving $W'$ is anyway unclear since  
the comparison between the observed $W'$ for the globular clusters and 
the $W'$ predicted by applying Eq.~(\ref{eq:wpfeh_fnxscl}) -- therefore 
neglecting the effect of [Ca/H] -- improves slightly at the low $W'$ end while 
gets slightly worse at intermediate $W'$. 

This analysis is clearly not exhaustive, but indicative of the effects
of the obvious sources of uncertainty \citep[see also][]{cenarro2002}. 
Since we are interested in understanding the effect on the [Fe/H]
determination, it is important to notice that notwithstanding the
large difference in [Ca/Fe] between calibrating globular clusters and
dSphs at the high [Fe/H] end ($\sim$ 0.5 dex at [Fe/H]$\sim -0.7$) and
the large difference in age (Fornax stars at $\sim -0.7$ have an age
of 3-6 Gyr) the error in estimating [Fe/H] using globular clusters as
calibrators is just 0.1-0.2 dex.

We point out that in this respect the Fornax dSph is one of the most
extreme cases as the large majority of dSphs do not show such an extended star
formation history, and so the difference in [Ca/Fe] (and in age) with
respect to the calibrating globular clusters will typically not be as
extreme.

\subsection{On age effects}
As shown in \citet{pont2004}, for metallicities $< -0.5$ dex, a
relative age difference of $\sim$12 Gyr with respect to calibrating
globular clusters results in an underestimate in metallicity of
$\sim$0.4 dex (their fig.~16, middle panel). This error decreases
substantially for smaller age differences. For instance, for a
metallicity $\sim -0.5$, the error induced by using a calibrating
cluster 12.6 Gyr old for a star which is 4 Gyr old is $\sim \pm $ 0.2
dex. At lower metallicities this effect is smaller.  The stars in the
Sculptor dSph have ages $>$ 8 Gyr \citep[e.g.,][]{mon1999} and
[Fe/H]$\la -1$, thus the age effect will be negligible. The bulk of
stars in the Fornax dSph have an age between 3-6 Gyr old
\citep[e.g.,][]{pont2004, battaglia2006}; only a small fraction of
stars have younger ages\footnote{these spectroscopic samples of RGB
stars only contain stars older than 1-2 Gyr}, thus the age effect
could result in an underestimate of $\sim$0.1-0.15 dex from the CaT
measurement for 3-6 Gyr old stars.  As mentioned above, most of the
dSphs do not show star formation histories as extended as Fornax, thus
the age effect should typically be even smaller.

\subsection{Synthetic stellar atmospheres}
Another way of approaching these issues is to model the expected behaviour of the
CaT lines using stellar atmospheres and synthetic spectra. It is
perhaps not surprising that these lines exhibit a
strong dependence on global metallicity, best represented by [Fe/H], rather than on the abundance
of calcium itself, given the strength of the features. With EWs
ranging from  $\sim$500m\AA\, to several \AA\,, the CaT lines are not only
saturated but are truly in the strong line regime of line
formation, when a significant part of the flux is found in the 
pressure-broadened wings rather than the core of the line. This effectively 
means that electronic pressure ($P_e$) is driving the strength of the line 
rather than the usual combination of temperature and elemental abundance that
shape weak lines. This can explain a fundamental
characteristic of the CaT: since many metals contribute to $P_e$, 
the CaT lines become sensitive to the global metallicity, [Fe/H], rather than
calcium abundance alone, through the pressure-dependent wings.
In fact, natural broadening dominates the wings of CaT lines, so
that line strength will increase with decreasing electronic
pressure (roughly as $1/P_e$). This explains why the CaT line strengths increase
with increasing luminosity (i.e. decreasing gravity and hence pressure).
The metallicity dependence of $P_e$ will therefore also contribute to shaping  
the CaT lines. Finally, global metallicity also plays a role in changing the
blanketing properties by numerous small metallic absorption lines, both in
the wavelength regions used to define the continuum and in the CaT
wings themselves. 
It is therefore difficult to speculate theoretically
how the CaT $W'$ should behave upon varying each stellar
parameter and we have therefore used synthetic stellar spectra, 
``observed'' with an approach mimicking our observational procedure.

\begin{figure*}
\begin{center}
\includegraphics[width=130mm]{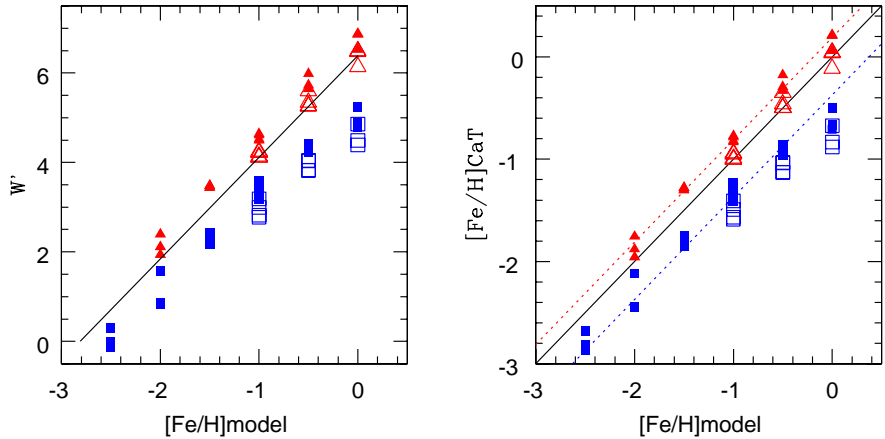}
\caption{Left: Relation between the $W'$ and 
the input metallicity of the synthetic spectra of the \citet{munari2005} 
grid ([Fe/H]model) along a 12Gyr isochrone. Filled squares
are the models with [$\alpha$/Fe]$=$0.0 and filled triangles are the
models with [$\alpha$/Fe]$=$0.4. The thick black line shows our
calibration. The open symbols show the effect of changing the age of
the isochrone from 12 to 2.5Gyrs for metallicities [Fe/H]$\geq-$1.
Right: Comparison between the metallicity obtained using our CaT
calibration on the synthetic spectra ([Fe/H]CaT) and the
input metallicity of the model ([Fe/H]model). Dotted lines show the
mean offsets of the [$\alpha$/Fe]$=$0.0 models and [$\alpha$/Fe]$=$0.4
models with respect to the one-to-one relation (solid line).
}
\label{fig:munari}
\end{center}
\end{figure*} 

We have checked the validity of calibrating [Fe/H] to the width of the CaT lines 
by using synthetic spectra of RGB stars. \citet{munari2005}
published a large grid of synthetic spectra covering the CaT region at
various resolutions.  Since our FLAMES LR resolution of $R \approx 6500$
did not correspond to the available resolutions we 
rebinned the Munari $R = 20000$ resolution spectra to the FLAMES LR
resolution.  The model atmosphere spectra cover a range
of stellar parameters including log$g$, $T_{\rm eff}$, [M/H] and also include
models computed with [$\alpha$/Fe]=$0.0$ and $+0.4$. The CaT EWs 
of a representative sample of model atmosphere spectra covering
the range of stellar RGB parameters encountered in dSphs were measured
in a similar manner to those of our observed LR spectra.

In particular, we measured the EW of a set of
synthetic stars taken along the upper RGB of a series of 12Gyr isochrones with
$-2.5 \leq$ [Fe/H] $\leq +0.0$ \citep{bertelli1994}. We then derived the corresponding [Fe/H] using our CaT
calibration based on globular clusters (Eq.~\ref{eq:feh_b07}). 
The $V$ magnitude for each synthetic star was read off the isochrone, while the 
$V_{\rm HB}$ was assumed [Fe/H] dependent, of the form 
$V_{\rm HB}=1.17 +0.39\times$[Fe/H] \citep[and references therein, which also fit
well the same isochrones]{ortolani1995}. The resulting ($V-V_{\rm HB}$) range from 0.5 to 3.5
(RGB tip), and are very similar to the range of luminosities of our targets in
Sculptor. 

The CaT EWs were measured in both $\alpha$-poor and
$\alpha$-rich models, and Fig.~\ref{fig:munari} shows how remarkably
well the  resulting metallicities from the CaT calibration ([Fe/H]CaT)
agree with the input model metallicity ([Fe/H]model).
Our calibration on the $\alpha$-rich synthetic spectra overestimates
the [Fe/H]model by only +0.19~dex with a negligible dispersion and no
particular trend, while our [Fe/H]CaT of the $\alpha$-poor spectra
underestimates [Fe/H]model by $-0.38$~dex in the mean. This is 
expected, as our calibration is based on globular clusters, which have
in the mean [$\alpha$/Fe]$\sim +0.25$, and are therefore closer to the
$\alpha$-rich grid of \citet{munari2005}.  Moreover, the relation between 
the CaT $W'$ and [Fe/H]model is linear, with a slope 
(for $\alpha$-rich) matching exactly the slope of our calibration.  This 
reassuring result supports the soundness of using simple CaT (linear) 
calibrations to derive metallicities.

We also examined the effect of using the same calibration for
synthetic stars along  younger RGB isochrones of 2.5Gyrs (similar to
the mean age of the stars in our Fornax sample). The
main effect is the luminosity increase: at a given metallicity, the
RGB isochrones of intermediate and old ages overlap almost perfectly
in temperature and gravity, but younger RGB stars appear more
luminous than old stars of the same temperature and gravity (because 
a younger star with the same gravity and temperature will be of 
higher mass, and hence more luminous). Between a 12 and 2.5\,Gyrs
isochrone, the luminosity at a given gravity increases by $\sim$0.5 mag, 
translating into a $-0.2$dex metallicity decrease once the CaT calibration is 
applied, as illustrated in Fig.~\ref{fig:munari}.  This age effect is small,
and comparable to the shallow slope observed in Fig.~\ref{fig:met_newcal} 
(lower panel) where our younger metal-rich stars (Fornax) tend to give lower 
[Fe/H]CaT abundances than expected.

As a final comment on this comparison, we would like to stress
that the CaT behaviour as a function of luminosity and metallicity
assumed in this work is fully consistent with what is expected from
synthetic spectra.  The synthetic spectra are however probably not precise 
enough to further constrain the CaT behaviour, as the core of these very strong
lines is almost certainly not well modelled by the stellar atmosphere
models, as the cores of the lines form close to the stellar surface, a
region which is problematic to model, and may even include a possible contribution from the
stellar chromosphere. 

\section{Summary and conclusions}
We described the data reduction steps we use within the DART collaboration to estimate velocities and 
CaT metallicities for LR data for RGB stars in dSphs. We showed that we obtain accurate velocities and [Fe/H] 
measurements, with internal error in velocity $\sim$ 2 \kms and in [Fe/H] $\sim$ 0.1 dex at a S/N per \AA\, of 20. 
 
We used 4 Galactic globular clusters observed with VLT/FLAMES in the CaT region
 to test the performance of several CaT $W'$-[Fe/H] relations existing 
in the literature. We also derived the best calibration from these globular cluster data. 
The relation here derived is consistent at the 1-$\sigma$ level with the calibration derived in 
\citet{tolstoy2001}, which we used 
in \citet{tolstoy2004}, \citet{battaglia2006} and \citet{helmi2006}. 

We used a sample of 93 and 36 RGB stars in the Sculptor and Fornax dSphs, respectively, overlapping between our LR 
and HR VLT/FLAMES 
observations, to test the globular clusters CaT calibration. This is the first time that the CaT calibration 
is tested on field stars in galaxies. We find a good agreement between the metallicities 
derived with these two methods. However, a systematic trend is present with [Fe/H], such that 
using the globular cluster calibration derived in this work 
the [Fe/H] measurement from CaT is overestimated of $\sim$0.1 dex at [Fe/H]$_{\rm HR} < -2.2$, 
whilst at [Fe/H]$_{\rm HR} > -1.2$ it is underestimated of $\sim$0.1-0.2 dex. No clear
systematic trend is instead derived from our data for [Fe/H]$_{\rm HR} > -0.8$. 
In order to understand this systematic effect, we explored the possible contribution of Ca abundance to the calibration, and 
showed that there are indications that it might well affect the CaT $W'$, although 
much less than [Fe/H]. From our dataset we also show that, contrary to previous claims, it is not advisable 
to use the CaT $W'$ as a linear indicator of [Ca/H]. 

Finally we investigated the effect of varying stellar atmosphere parameters on the CaT method by
analysing a large sample of model atmosphere spectra \citep{munari2005}.   We again demonstrated
that the CaT method is (surprisingly) robust to the usual combination of age, metallicities and
[$\alpha$/Fe] variations seen in nearby dSphs.

From our analysis we see that even for 
large differences in [Ca/Fe] between calibrating globular clusters and our sample of dSphs 
($\sim$ 0.5 dex at [Fe/H]$\sim -0.7$) and large difference in ages 
(at [Fe/H] $\sim -0.7$ Fornax stars are $\sim$ 10 Gyr younger than globular clusters stars) the error in 
estimating [Fe/H] using globular clusters as calibrators is just 0.1-0.2 dex. 
The Fornax dSph is likely to represent the most 
extreme case as it has had one of the most extended star formation histories among the dSphs in the Local Group. 

We conclude that CaT-[Fe/H] relations calibrated on globular clusters can be applied with confidence 
to RGB stars in composite stellar populations such as galaxies, at least in the [Fe/H] 
range probed by the above analyses, $-2.5 <$ [Fe/H] $<-0.5$. Hence, the CaT method provides a 
good indicator of the overall metallicity of resolved stars.

This has implications for the efficiency with which we can obtain metallicity distribution 
functions of nearby resolved galaxies. The HR data collected in this paper required more than 6 nights 
of VLT observing time for $\sim$ 150 spectra, whereas $\sim 120$ CaT spectra were obtained in one hour VLT observing time. 

\section*{Acknowledgements}
We acknowledge F.Pont for kindly providing the equivalent widths of the individual CaT lines in electronic form. 
We thank A.Cole for useful discussions. AH acknowledges support from NWO and NOVA.

\bibliographystyle{mn}
\bibliography{mn-jour,gcs_biblio}

\begin{table*}
\begin{center}
\begin{tabular}{lccccccccc}
\hline
 Cluster  &  RA(J2000) & DEC(J2000) & V$_{\rm HB}$ & v$_{\rm sys}$ [\kms]& [Fe/H]$_{\rm CG97}$ & [Fe/H]$_{\rm R97}$ & [Fe/H]$_{\rm T01}$ & [Fe/H]$_{\rm C04}$ & [Ca/Fe]\\
\hline
 NGC104  &  $00^h 24^m 05.2^s$ & $-72^{\circ} 04' 51''$ & 14.06 & $-$14.4$\pm$0.7 &$-$0.70$\pm$0.03 &  $-$0.78$\pm$0.05 & $-$0.68$\pm$0.06  & $-$0.82$\pm$0.06 & 0.30\\
 NGC5904 &  $15^h 18^m 33.8^s$ & $+02^{\circ} 04' 58''$ & 15.07 & 53.2$\pm$0.7  &$-$1.11$\pm$0.03 &  $-$1.13$\pm$0.07 & $-$1.02$\pm$0.08  & $-$1.19$\pm$0.06 & 0.21\\
 NGC3201 &  $10^h 17^m 36.8^s$ & $-46^{\circ} 24' 40''$ & 14.77 & 497.5$\pm$1.2 &$-$1.23$\pm$0.05 &  $-$1.27$\pm$0.07 & $-$1.17$\pm$0.06  & $-$1.36$\pm$0.07 & 0.11 \\
 NGC4590 &  $12^h 39^m 28.0^s$ & $-26^{\circ} 44' 34''$ & 15.68 & $-$98.3$\pm$1.7 &$-$1.99$\pm$0.06 &  $-$1.94$\pm$0.06 & $-$1.90$\pm$0.07  & $-$2.11$\pm$0.06 & 0.32 \\
\hline
\end{tabular}
\caption{The globular clusters used in this work. Positions and V$_{\rm HB}$ 
are from \citet{harris1996}. The systemic velocity v$_{\rm sys}$ was derived in this work. 
[Fe/H]$_{\rm CG97}$ is the value of metallicity from \citet{carretta1997}. 
[Fe/H]$_{\rm R97}$, [Fe/H]$_{\rm T01}$, [Fe/H]$_{\rm C04}$ are the values of the metallicity from this work, 
using the calibrations from \citet{rutledge1997a, rutledge1997b}, 
\citet{tolstoy2001}, \citet{cole2004}, i.e. 
Eqs.~(\ref{eq:feh_r97}),(\ref{eq:feh_t01}),(\ref{eq:feh_c04}). The errors in 
the derived metallicities were obtained from the standard deviation of the values of reduced equivalent 
width for each globular cluster. The [Ca/Fe] values are from \citet{carretta2004} for 47 Tuc and all the other ones 
are listed in \citet{koch2006}.
} \label{tab:gc}
\end{center}
\end{table*}

\begin{table*}
\begin{center}
\begin{tabular}{lccccc}
\hline
 Object  &  RA(J2000) & DEC(J2000) & Date of observation [YY-MM-DD] & exptime [s] & grating \\
\hline
Sculptor &  01 00 03 & -33 41 30   &  2003-08-20  &   4200        &  HR13 \\
Sculptor &  01 00 03 & -33 41 30   &  2003-08-21  &   3x3600, 4700  &  HR14\\
Sculptor &  01 00 03 & -33 41 30   &  2003-08-22  &   2x3600  & HR13 \\
Sculptor &  01 00 03 & -33 41 30   &  2003-08-22  &   4500  & HR14 \\
Sculptor &  01 00 03 & -33 41 30   &  2003-08-23  &   5400  & HR14\\
Sculptor &   01 00 03 & -33 41 30   &  2003-08-24  &   2x3600  & HR10\\
Sculptor &  01 00 03 & -33 41 30   &  2003-08-25  &   5400  & HR10\\
Sculptor &  01 00 03 & -33 41 30   &  2003-08-28  &   3600  & HR10\\
Sculptor &  00 58 22 & -33 42 36   &  2003-09-29  &   2x1800  & LR8 \\
Sculptor &  01 01 44 & -33 43 01   &  2003-10-01  &   2x1800  & LR8 \\
Sculptor &  01 00 03 & -33 41 29   &      2004-09-10  &                3600  &     LR8\\ 
Sculptor &  01 01 11 & -33 56 50 &        2005-11-07&     3600  &  LR8 \\
Sculptor &  00 58 59 & -33 27 04 &        2005-11-08 &    2x2700 &  LR8 \\  
Fornax &           02 39 30 & -34 28 39 &    2003-09-29 &    2x3600   & HR13 \\
Fornax &           02 39 30 & -34 28 39 &    2003-09-29 &    2x3100  & HR14 \\
Fornax &           02 39 30 & -34 28 39 &    2003-09-29 &    2x3600   & HR13 \\
Fornax &           02 39 30 & -34 28 39 &    2003-09-30 &    3300, 2x5400 &  HR14\\
Fornax &           02 39 30 & -34 28 39 &    2003-09-30 &    3600   & HR10 \\
Fornax &           02 39 30 & -34 28 39 &    2003-10-01 &    3x3600   & HR10 \\  
Fornax &           02 39 30 & -34 28 39 &    2003-10-01 &    3300, 3600 &  HR14\\
Fornax &           02 39 30 & -34 28 39 &    2004-01-14 &    3600   & HR10 \\
Fornax &           02 39 30 & -34 28 39 &    2004-01-15 &    3600   & HR13 \\
Fornax &           02 39 30 & -34 28 39 &    2004-01-19 &    2x3600   & HR13 \\
Fornax &           02 39 30 & -34 28 39 &    2004-01-20 &    3600   & HR13 \\
Fornax &           02 39 30 & -34 28 39 &    2004-01-21 &    2x3600  &  HR14  \\
Fornax &           02 39 30 & -34 28 39 &    2004-01-22 &    3600,3900 &  HR14 \\
Fornax &           02 39 30 & -34 28 39 &    2004-01-23 &    3600   & HR10 \\
Fornax &           02 39 30 & -34 28 39 &    2004-01-23 &    3600  & HR14  \\  
Fornax &           02 39 30 & -34 28 39 &    2004-01-24 &    3600   & HR10 \\  
Fornax &           02 39 30 & -34 28 39 &    2004-01-26 &    3600   & HR10 \\  
Fornax &    02 38 56 & -34 24 48      &     2005-11-07    &  3600 &   LR8   \\
Fornax &           02 39 36 & -34 40 51  &            2003-08-24 &    1500,1800 &  LR8 \\
Fornax &           02 39 58 & -34 15 28  &           2003-08-26  &   2400  &   LR8 \\
NGC104    &      00 22 15 & -72 04 45    &          2005-01-11     &         3x600     &    LR8 \\
NGC3201   &      10 17 36 & -46 24 40    &          2003-02-22     &         300, 327  & LR8 \\
NGC3201   &      10 17 36 & -46 24 40    &          2003-03-06     &         2x600  & LR8 \\
NGC4590   &      12 39 28 & -26 44 33    &          2003-03-06     &         3x600  & LR8 \\
NGC5904   &      15 18 03 &  02 06 40    &          2003-03-05     &         6x600  & LR8 \\
\hline         
\end{tabular}
\caption{Table of VLT/FLAMES observations for the Sculptor and Fornax dSphs and the globular 
clusters NGC104, NGC3201, NGC4590, NGC5904. We used these observations for the analysis of the 
CaT-[Fe/H] calibration.} \label{tab:journal}
\end{center}
\end{table*}

\newpage

\begin{table*} 
\tabcolsep 1.3pt
\begin{center}
\begin{tabular}{llcccccccccc}
\hline
 Object & Star ID  &  RA(J2000) & DEC(J2000) & $V$ & e$V$ & $v_{\rm hel}$ & e$v_{\rm hel}$ & $\Sigma W_{\rm T01}$ & e$\Sigma W_{\rm T01}$ & 
[Fe/H]$_{\rm LR}$ & e[Fe/H]$_{\rm LR}$ \\
\hline
NGC104   & GB07-TUC1002        &  0 23 41.71 &  $-$72  5  2.6 &     13.052 &     0.002 &    $-$8.55 &     0.38 &   5.209 &   0.055 &   $-$0.80 &    0.02 \\
NGC104   & GB07-TUC1510        &  0 24 25.01 &  $-$72  2 45.7 &     12.652 &     0.001 &    $-$6.93 &     0.57 &   5.729 &   0.039 &   $-$0.69 &    0.02 \\
NGC3201  & GB07-N3201-S60      & 10 17 42.84 &  $-$46 22 48.5 &     13.601 &     0.002 &   495.55 &     0.65 &   4.144 &   0.055 &   $-$1.32 &    0.02 \\
NGC3201  & GB07-N3201-S74      & 10 17 48.53 &  $-$46 23 39.3 &     14.602 &     0.002 &   498.98 &     0.63 &   3.636 &   0.061 &   $-$1.26 &    0.03 \\
NGC4590  & GB07-M68-S99        & 12 39 40.21 &  $-$26 47 37.5 &     15.186 &     0.004 &  $-$100.62 &     1.80 &   2.021 &   0.132 &   $-$2.06 &    0.06 \\
NGC4590  & GB07-M68-S144       & 12 39 32.06 &  $-$26 47 53.6 &     15.291 &     0.004 &   $-$99.36 &     1.92 &   2.139 &   0.139 &   $-$1.98 &    0.06 \\
NGC5904  & GB07-M5-S23         & 15 17 55.26 &    2  7 25.6 &     14.544 &     0.002 &    55.11 &     1.08 &   4.175 &   0.036 &   $-$1.12 &    0.02 \\
NGC5904  & GB07-M5-S980        & 15 18 16.42 &    2  7 18.5 &     14.264 &     0.002 &    46.88 &     1.16 &   4.525 &   0.023 &   $-$1.05 &    0.01 \\
\hline
\end{tabular}
\caption{This table lists the relevant data for the stars in the 
globular clusters NGC104, NGC3201, NGC4590, NGC5904 observed with VLT/FLAMES at LR, and that we used in the analysis 
of the CaT-[Fe/H] calibration. For the CaT analysis, we select only those stars whose colour and magnitude are consistent with what
is expected for RGB stars, and that have $V-V_{\rm HB} <$ 0, S/N$>$ 10/\AA\, error in velocity $<$ 5 \kms and velocities 
consistent with membership to the cluster (we assign membership to those stars within 3$\sigma$ of the systemic velocity, see Table~\ref{tab:gc}). 
The columns indicates: (1) the globular cluster name; (2) the star ID; 
(3),(4) star coordinates (right ascension in hours and declination in degrees); (5),(6) $V$ magnitude and its error; 
(7),(8) heliocentric velocity and its error; (9),(10) summed CaT equivalent width ($EW_2 + EW_3$) and its error; 
(11),(12) [Fe/H] value and its error, derived from the LR observations applying Eq.~(\ref{eq:feh_b07}). The photometry and astrometry
are from \citet{stetson2000}. } \label{tab:results_gcs}
\end{center}
\end{table*}

\begin{landscape}
\begin{table} 
\tabcolsep 1.3pt
\begin{center}
\begin{tabular}{llcccccccccccccc}
\hline
 Object & Star ID  &  RA(J2000) & DEC(J2000) & $V$ & e$V$ & $v_{\rm hel}$ & e$v_{\rm hel}$ & $\Sigma W_{\rm T01}$ & e$\Sigma W_{\rm T01}$ & 
[Fe/H]$_{\rm LR}$ & e[Fe/H]$_{\rm LR}$ & [Fe/H]$_{\rm HR}$ & e[Fe/H]$_{\rm HR}$ & [Ca/Fe]$_{\rm HR}$ & e[Ca/Fe]$_{\rm HR}$\\
\hline
Sculptor & GB07-Scl01 &  1  0  2.69 &  -33 30 25.3 &     18.017 &     0.003 &   109.24 &     1.31 &   4.269 &   0.250 &   -1.53 &    0.11 &   -1.39 &    0.10 &    0.10 &    0.12 \\
Sculptor & GB07-Scl02 &  1  0 25.78 &  -33 30 25.4 &     17.692 &     0.002 &   119.62 &     1.49 &   3.392 &   0.267 &   -2.00 &    0.12 &   -1.93 &    0.11 &    0.30 &    0.13 \\
Sculptor & GB07-Scl03 &  0 59 37.00 &  -33 30 28.4 &     17.706 &     0.002 &   101.70 &     1.01 &   4.344 &   0.087 &   -1.58 &    0.04 &   -1.63 &    0.10 &    0.27 &    0.13 \\
Sculptor & GB07-Scl04 &  0 59 55.63 &  -33 33 24.6 &     16.890 &     0.002 &   120.51 &     1.13 &   3.793 &   0.121 &   -2.05 &    0.05 &   -1.88 &    0.10 &    0.23 &    0.13 \\
Sculptor & GB07-Scl05 &  0 59 47.21 &  -33 33 36.9 &     17.608 &     0.002 &   105.13 &     0.85 &   3.362 &   0.162 &   -2.04 &    0.07 &   -2.25 &    0.10 &    0.37 &    0.13 \\
Sculptor & GB07-Scl06 &  1  0 21.08 &  -33 33 46.4 &     18.082 &     0.003 &   107.34 &     1.94 &   4.755 &   0.221 &   -1.30 &    0.10 &   -1.30 &    0.10 &    0.17 &    0.13 \\
Fornax   & GB07-Fnx01 &  2 39 22.15 &  -34 19 40.4 &     18.355 &     0.004 &    68.45 &     0.88 &   4.155 &   0.201 &   -1.81 &    0.09 &   -1.44 &    0.05 &    0.02 &    0.10 \\
Fornax   & GB07-Fnx02 &  2 39 51.42 &  -34 21 21.0 &     18.548 &     0.005 &    59.97 &     2.08 &   5.869 &   0.217 &   -1.00 &    0.10 &   -0.79 &    0.04 &   -0.44 &    0.05 \\
Fornax   & GB07-Fnx03 &  2 39 31.49 &  -34 23  5.1 &     18.534 &     0.005 &    53.72 &     0.88 &   5.564 &   0.150 &   -1.14 &    0.07 &   -0.85 &    0.05 &   -0.17 &    0.07 \\
Fornax   & GB07-Fnx04 &  2 38 49.28 &  -34 24  5.0 &     18.543 &     0.004 &    55.20 &     1.61 &   6.097 &   0.145 &   -0.90 &    0.06 &   -0.93 &    0.06 &   -0.22 &    0.07 \\
Fornax   & GB07-Fnx05 &  2 39  4.31 &  -34 25 18.9 &     18.390 &     0.004 &    55.64 &     1.89 &   5.275 &   0.154 &   -1.31 &    0.07 &   -0.95 &    0.04 &   -0.11 &    0.06 \\
Fornax   & GB07-Fnx06 &  2 38 55.53 &  -34 25 36.3 &     18.188 &     0.004 &    91.68 &     1.07 &   2.945 &   0.135 &   -2.39 &    0.06 &   -2.65 &    0.04 &    0.36 &    0.06 \\
\hline
\end{tabular}
\caption{This table lists the relevant data 
for the stars observed with VLT/FLAMES at both LR and HR resolution in the Sculptor and Fornax dSphs, and used in the analysis 
of the CaT-[Fe/H] calibration. The columns indicates: (1) the galaxy name; (2) the star ID; 
(3),(4) star coordinates (right ascension in hours and declination in degrees); (5),(6) $V$ magnitude and its error; 
(7),(8) heliocentric velocity and its error; (9),(10) summed CaT equivalent width ($EW_2 + EW_3$) and its error; 
(11),(12) [Fe/H] value derived from the LR observations applying Eq.~(\ref{eq:feh_b07}) and its error; (13),(14) HR [Fe/H] value and error; 
(14), (15) [Ca/Fe] and its error, from the HR observations. The photometry and astrometry
are from our ESO/WFI observations \citep{tolstoy2004, battaglia2006}. 
Those stars with [Ca/Fe]$=-9.99$ had too low signal-to-noise to allow a determination of the HR [Ca/H].
} \label{tab:results_dsphs}
\end{center}
\end{table}
\end{landscape}

\label{lastpage}

\end{document}